\documentclass[aps,pra,showpacs,twocolumn,floatfix,nofootinbib]{revtex4}
\usepackage{amsmath}    % need for subequations
\usepackage{graphicx}   % need for figures
\usepackage{verbatim}   % useful for program listings
\usepackage{color}      % use if color is used in text
\usepackage{subfigure}  % use for side-by-side figures
\usepackage{hyperref}   % use for hypertext links, including those to external documents and URLs
\usepackage[bbgreekl]{mathbbol}
\usepackage{amsfonts}
\newcommand{\vct}[1]{{\bf #1}}
\usepackage{subfigure}
\renewcommand\Re{\operatorname{Re}}
\renewcommand\Im{\operatorname{Im}}

\begin{document}
\title{Heat radiation from long cylindrical objects}
%\numberwithin{equation}{section}

\author{Vladyslav~A. Golyk}
\affiliation{Massachusetts Institute of Technology, Department of
  Physics, Cambridge, Massachusetts 02139, USA}
\author{Matthias Kr\"uger}
\affiliation{Massachusetts Institute of Technology, Department of
  Physics, Cambridge, Massachusetts 02139, USA}
\author{Mehran Kardar}
\affiliation{Massachusetts Institute of Technology, Department of
 Physics, Cambridge, Massachusetts 02139, USA}

\begin{abstract}
The heat radiated by objects small or comparable
to the thermal wavelength can be very different from the classical blackbody radiation
 as described by the laws of  Planck and Stefan-Boltzmann. We use methods
based on scattering of electromagnetic waves  to explore
the dependence on size, shape, as well as material properties. In
particular, we explore the radiation from a long cylinder at uniform temperature, discussing in detail the degree of polarization of the emitted radiation.
If the radius of the cylinder is much smaller than the thermal
wavelength, the radiation is polarized parallel to the cylindrical axis and becomes perpendicular when the radius is comparable to the thermal wavelength. For a cylinder of uniaxial material (a simple model for carbon nanontubes), we find that the influence of uniaxiality on the polarization is most pronounced if the radius  is larger than a few microns, and quite small for submicron sizes typical for nanotubes.
\end{abstract}


\maketitle

\section{Introduction}

Thermal radiation lies at the heart of modern statistical physics and goes back to the beginnings of quantum mechanics more than a century ago \cite{Planck}. Planck's law describes the intensity $\mathcal{I}$ (per unit surface area and solid angle) of radiation of a black body at temperature $T$
\begin{equation}
\mathcal{I}=\frac{\hbar \omega^3}{2\pi^2c^2}\frac{1}{e^{\frac{\hbar\omega}{k_BT}}-1}.\label{eq:Planck}
\end{equation}
Here, $\omega$ is the  {angular} frequency of radiation and $\hbar$ and $c$ are  {reduced} Planck's constant and the speed of light, respectively.  Integration over angles and frequencies yields the well known Stefan-Boltzmann law\cite{Boltzmann} for the total power $H$ radiated per unit area $A$,
\begin{equation}
H/A=\sigma T^4,\label{SB}
\end{equation}
with $\sigma= \pi^2 k_B^4 / (60 \hbar^3 c^2)$.

Only recently, various phenomena, leading to modifications of these laws have been explored. For example, the effect of spatial coherence of emitted heat radiation %by planar surfaces
was studied by many authors \cite{Carminati,Barabanenkov,Laroche,Dahan} as this effect can be used in many technological applications \cite{Laroche2,Cvitkovic,Soljacic,Biener}, such as thermophotovoltaic and high-efficiency light sources.

For real materials, Eq.~\eqref{eq:Planck} can be adjusted by introducing the (angle dependent) emissivity of the material, which is unity for a black body.
Considering objects with sizes smaller or comparable to the thermal wavelength $\lambda_T=\hbar c/k_BT$ (approximately 7.6 $\mu$m at room temperature), the radiated energy differs from the above equations because of interference effects of the object with the emitted radiation. In other words, the emissivity then depends on the size and shape of the object. Additionally, if the object is smaller than the penetration (skin) depth, the emitted power is proportional to the object's volume, rather than surface area. The heat radiation of small spherical objects including these effects have been studied since the 1970s\cite{Bohren,Kattawar70}. Also, effects of excitations\cite{Hansen} and electric currents\cite{Shapiro} on the radiation have been studied.  Furthermore, recent studies on superscattering properties of subwavelength nanostructures (e.g. nanorods)\cite{Fan} make such systems potential candidates for efficient heat transfer sources.

Experimentally, the radiation of thin cylindrical objects with thickness in the range of the thermal wavelength is very well accessible, and has e.g. been studied using metal wires with interesting findings: 50 years ago, it was discovered that the radiation of a hot
thin metal wire is significantly polarized \cite{Ohman}. Polarizations of 28\%\cite{Ohman} and
50\%\cite{Agdur} %percent
in the direction perpendicular to the wire were
measured for thin incandescent tungsten and silver wires
respectively. In both studies the thickness of the wires was larger or
comparable to the thermal wavelengths. These findings triggered a number of
studies on the properties of thermal radiation
of sources of various designs, including platinum microwires\cite{Ingvarsson,
Skulason}, semiconductor layers in external magnetic
fields\cite{Kollyukh}, bundles of carbon nanotubes\cite{Li, Aliev} and SiC lamellar gratings\cite{Marquier}.

For wires with thickness smaller or comparable to the
thermal wavelength, the radiation was found (e.g. for platinum) to be polarized in the direction
parallel to the wire
becoming fully polarized as the width
approaches zero~\cite{Ingvarsson,Skulason}. Polarization effects
have also been observed for radiation of bundles of carbon
nanotubes~\cite{Li} and are considered a simple way of finding the
degree of alignment inside the bundle. For carbon nanotubes, an explanation for this polarization, taking into account the electronic structure of the tubes, has  been discussed in Ref.~\cite{Aliev}.

Recent work on the heat radiation of thin metal wires \cite{bimonte1} provides experimental as well as theoretical results, albeit restricted to emission perpendicular to the cylindrical axis. Also, a series of works~\cite{regan1,regan2,regan3} discuss radiation emitted by individual incandescent carbon nanotubes. In Ref.\cite{regan2}, the polarization of the radiation is studied both experimentally and theoretically using a model based on Mie theory.

 {In this paper, we provide the general formalism to compute the heat radiation of arbitrary objects in terms of their classical scattering properties, which is part of the general framework~\cite{Kruger} for non-equilibrium electromagnetic fluctuations involving multiple objects and arrays~\cite{bimonte2, Antezza, Coley, johnson, Alex, sphereplate, bimonte3}. Thereby we give a more extensive derivation of the corresponding results presented in Ref.\cite{Kruger}, which is also more general in terms of material properties: We include the  possibility of dielectric or magnetic losses, locality or non-locality. Additionally, we provide a new, simpler, formalism to derive the heat radiation (see Eq.~\eqref{eq:new} below).} To this end, we start from quantum thermal fluctuations inside the object following the theory of fluctuational electrodynamics introduced over 60 years ago by Rytov~\cite{Rytov}. Then, we derive the heat radiation of the experimentally important case of a cylindrical object, discussing polarization effects for different conducting and insulating materials,  {as well as asymptotic limits. This detailed dicussion hence provides much deeper understanding of material and size effects compared to the brief introduction of dielectric cylinders in Ref.~\cite{Kruger}}. We derive the  {so far unknown} heat radiation of a cylinder made of uniaxial material, where the symmetry axis of the material and the cylinder coincide. We apply this formula to introduce a simple model for the heat radiation of multi-walled carbon nanotubes (MWCNT).

The paper is composed as follows:  {In section \ref{section2}, we give the general formalism for heat radiation of arbitrary objects and apply it to the cases of a cylindrical object. Additionally, we find the radiation of an anisotropic plate, generalizing the known results \cite{Rytov} for the isotropic case}.  Sec. \ref{section3} finally gives the specific form of the scattering operator for a cylinder made of uniaxial material, needed to evaluate the formula of Sec. \ref{section2.2}.  In Sec. \ref{section4}, we discuss the total radiated energy by cylinders made of different materials, such as dielectrics, metals, and MWCNT, putting special focus on the  degree of polarization of the radiation.  Finally, in Sec.~\ref{section5}, we discuss the spectral density of the energy emitted by these materials, a quantity which might be most accessible in experiments.  We close with a summary and discussion of our findings in Sec.~\ref{section6}.

\section{Heat radiation in terms of scattering
operator}\label{section2}
\subsection{General formalism for arbitrary objects} \label{section2.1}
Consider an  object with  homogeneous temperature $T_{obj}$ placed in
vacuum, enclosed by an environment (e.g. a cavity much larger than all other scales in the system) at temperature $T_{env}$. In global equilibrium, i.e.
 with $T_{obj}=T_{env}=T$, the autocorrelation function $C$
of the electric field is related to the imaginary part of the dyadic
Green's function $G_{ij}$ of the object by the
fluctuation-dissipation theorem (FDT)~\cite{Rytov, Eckhardt},
\begin{equation}\label{kruger1}
\begin{split}
C_{ij}^{eq}(T)&\equiv\left\langle
E_i(\omega;\textbf{r}) E_j^*(\omega;\textbf{r}')
\right\rangle^{eq}\equiv {\left\langle
\vct{E}(\omega;\textbf{r}) \otimes \vct{E}^*(\omega;\textbf{r}')
\right\rangle^{eq}_{ij}}\\
&=\left[a_T(\omega)+a_0(\omega)\right] \frac{c^2}{\omega^2}\textrm{Im}
G_{ij}(\omega;\textbf{r},\textbf{r}'),\
\end{split}
\end{equation}
 {where $\otimes$ denotes a dyadic product}. $a_T(\omega)\equiv \frac{\omega^4 \hbar
  (4\pi)^2}{c^4}(\exp[\hbar \omega/k_BT]-1)^{-1}$ describes the thermal contribution to quantum fluctuations, compare Eq.~\eqref{eq:Planck}.
 The zero point
fluctuations, which contribute $a_0(\omega)\equiv \frac{\omega^4
\hbar (4\pi)^2}{2c^4}$, are independent of the object's temperature and
 do not contribute to heat radiation. Hereafter we use the operator
notation $\mathbb{G}\equiv G_{ij}(\omega ;\textbf{r},\textbf{r}')$, where operator multiplication implies an integration over space as well as a $3\times 3$ spatial matrix multiplication, e.g. for the operators $\mathbb{A}$ and $\mathbb{B}$ (using Einstein summation convention),
\begin{equation}
(\mathbb{A}\mathbb{B})_{ik}(\vct{r},\vct{r}'')=\int d^3r' A_{ij} (\vct{r},\vct{r}') B_{jk}(\vct{r}',\vct{r}'').
\end{equation}
The Green's function is the solution of \cite{Jackson,Rahi}
\begin{equation}
\left[ \mathbb{H}_0+\mathbb{V}-\frac{\omega^2}{c^2}\mathbb{I}\right]\mathbb{G}=\mathbb{I},\label{eq:H}
%\left[ \boldsymbol{\nabla}\times\frac{1}{\mu(\vct{r})}\boldsymbol{\nabla}\times-\frac{\omega^2}{c^2}\widehat\varepsilon(\vct{r})\cdot\right]\mathbb{G}=\mathbb{I},\label{eq:H}
\end{equation}
which follows because the electric field obeys the Helmholtz equation, Eq.~\eqref{eq:H2} below. Here, $\mathbb{H}_0=\boldsymbol{\nabla}\times\boldsymbol{\nabla}\times$ describes free space, and $\mathbb{V}=\frac{\omega^2}{c^2}(\mathbb{I}-\bbespilon+\boldsymbol{\nabla}\times\left(\frac{1}{\bbmu}-\mathbb{I}\right)\boldsymbol{\nabla}\times)$ is the potential introduced by the object.
$\bbespilon$ and $\bbmu$ are the
complex (possibly nonlocal) dielectric permittivity and magnetic permeability tensors of the object. For isotropic and local materials, they reduce to scalars (e.g. $\bbespilon=\varepsilon\mathbb{I}$). $\mathbb{G}_{0}$ is
 the Green's function of free space. Using the identities
$\textrm{Im}
\mathbb{G}=-\mathbb{G}\textrm{Im} \mathbb{G}^{-1}\mathbb{G}^*$ and
$\textrm{Im} \mathbb{V}=
\textrm{Im}
(\mathbb{G}^{-1}-\mathbb{G}_{0}^{-1})$ \cite{Eckhardt}, which can be found from Eq.~\eqref{eq:H}, we
obtain
\begin{align}\label{kruger2}
C^{eq}(T)&=C_0+C(T)-a_T(\omega) \frac{c^2}{\omega^2}\mathbb{G}\textrm{Im} \mathbb{G}_0^{-1} \mathbb{G}^*,\\
C(T_{obj})&=-a_{T_{obj}}(\omega)\frac{c^2}{\omega^2}\mathbb{G}\notag\textrm{Im} \mathbb{V}\mathbb{G}^*\\&=-a_{T_{obj}}\frac{c^2}{\omega^2}(\omega) \int\limits_{\rm obj}d^3r' d^3r'' G_{ij}(\vct{r},\vct{r}')\notag\\&\times\textrm{Im} V_{jk}(\vct{r}',\vct{r}'') G_{kl}^*(\vct{r}'',\vct{r}'''),\label{eq:CT}
\end{align}
where $C_0=a_0(\omega)\frac{c^2}{\omega^2}\textrm{Im} \mathbb{G}$ is
the zero point term.  Equation~\eqref{kruger2} shows two different finite
temperature contributions to the electric field in equilibrium.
$C(T)$ contains an explicit integral over the sources within the
object, as $\Im \mathbb{V}$ is only nonzero inside the object, and we identify it with the desired heat radiation from the object. The expression in Eq.~\eqref{eq:CT} can be shown to be identical to expressions in the literature for both complex electric and magnetic permeabilities \cite{Eckhardt, Rytov}, where, in general, one has two terms, including $\Im \bbespilon$ and $\Im \bbmu$, respectively. The introduction of the potential $\mathbb{V}$ appears useful here, as it allows for a compact notation including both terms. The third term in Eq.~\eqref{kruger2},
\begin{equation}
C^{env}(T_{env})=-a_{T_{env}}(\omega)\frac{c^2}{\omega^2}\mathbb{G} \textrm{Im}
\mathbb{G}_0^{-1} \mathbb{G}^*,\label{eq:Cenv}
\end{equation}
is the contribution sourced
by
the environment.  {As a specific model for environment, consider the objects enclosed in a very large black cavity maintained at temperature $T_{env}$.} This latter identification can be corroborated on a different route by introducing a cold object
into the thermal background field $\mathbf{E}_0$ sourced by the environment, with  {field correlator given by} $\langle \vct{E}_0 \otimes \vct{E}_0^*\rangle=a_{T_{env}}(\omega)\frac{c^2}{\omega^2}\textrm{Im}\mathbb{G}_0$. At this point, it is useful to introduce the $\mathbb{T}$ operator or scattering amplitude $\mathbb{T}$\cite{Tsang, Rahi} of the object. It relates the homogeneous solution (also sometimes called the exciting field \cite{Tsang}) $\vct{E}_h$ (for $\mathbb{V}=0$) of the Helmholtz equation
\begin{equation}
\left[ \mathbb{H}_0+\mathbb{V}-\frac{\omega^2}{c^2}\mathbb{I}\right]\vct{E}=0,\label{eq:H2}
%\left[\boldsymbol{\nabla}\times\frac{1}{\mu(\vct{r})}\boldsymbol{\nabla}\times-\frac{\omega^2}{c^2}\widehat\varepsilon(\vct{r})\cdot\right]\vct{E}=0\label{eq:H2}
\end{equation}
to its (inhomogeneous) solution $\vct{E}_{ih}$ with the object present. This solution can be stated in terms of the Lippmann-Schwinger equation
\begin{equation}
\textbf{E}_{ih}=(1-\mathbb{G}_0\mathbb{T})\textbf{E}_h\label{eq:defT}.
\end{equation}
With this equation, the above introduction of the cold object into the free environment field is readily done and  $C^{env}$ is then the correlator of the field $\vct{E}_s$, the inhomogeneous solution with the object present,
\begin{align}\label{coldobject}
&C^{env}(T_{env})=\langle \vct{E}_s\otimes \vct{E}_{s}^*\rangle=(1-\mathbb{G}_0\mathbb{T})\langle \vct{E}_0\otimes\vct{E}^*_0\rangle\notag\\&
\times(-\mathbb{T}^*\mathbb{G}^*_0+1)
=-a_{T_{env}}(\omega)\frac{c^2}{\omega^2}\mathbb{G}\textrm{Im}
\mathbb{G}_0^{-1} \mathbb{G}^*,\
\end{align}
in agreement with Eq.~\eqref{eq:Cenv}. Here we used the identity \cite{Rahi}
\begin{equation}
\mathbb{G}=\mathbb{G}_0-\mathbb{G}_0\mathbb{T}\mathbb{G}_0\label{eq:GT}.
\end{equation}
Equation \eqref{coldobject} highlights the  {physical} interpretation of $C^{env}$:  {It is the radiation sourced by the environment and scattered by the object.}

Having found the contributions of the different sources (environment and object), one can now vary the temperature of these independently in order to arrive at the field outside the object when its temperature is different from that of the environment. If $T_{env}=0$ this field corresponds to the heat radiation of the object. To this end, we notice that it is not necessary to derive all the terms in Eq.~\eqref{kruger2}  {as explained in the following:} The explicit expression for $C(T_{obj})$ in Eq.~\eqref{eq:CT} contains the Green's function with one argument inside and one argument outside the object. While this function can be in principle derived, we find it more convenient to express $C(T_{obj})$ in terms of the Green's function with both arguments outside the object, as it is directly linked to the scattering operator by Eq.~\eqref{eq:GT}. Therefore, it is interesting to note that $C^{env}$ has all the sources outside the object and hence can be found in terms of this Green's function. While this is already obvious in Eq.~\eqref{coldobject} we additionally present a more rigorous way to derive $C^{env}$. The environment sources, described by  $\varepsilon_{env}$, can be thought of as being everywhere in the infinite space complementary to the object, infinitesimal in strength (environment ``dust'' \cite{Eckhardt}), i.e. $\varepsilon_{env}\to1$.
$C^{env}$ in Eq.~\eqref{eq:Cenv} can hence be written
\begin{align}
&C^{env}(T_{env})\notag=a_{T_{env}} (\omega)\\&\lim_{\varepsilon_{env}\to1}\int_{\rm outside}d^3
r'\widetilde{G}_{ik}(\textbf{r},\textbf{r}')\textrm
{Im}\varepsilon_{env}
\widetilde{G}^*_{jk}(\textbf{r}'',\textbf{r}'),\label{cenv}
\end{align}
 {which is identical to Eq.~(4) in  Ref. \cite{Kruger}}.Here, we introduced a Green's function $\tilde{\mathbb{G}}$ with $\mathbb{V}$ inside the object and
$\varepsilon_{env}$ outside. This is a simple modification of
$\mathbb{G}$ as a finite $\varepsilon_{env}-1$ only changes the
external speed of light so that $c$ in $\mathbb{G}$ is replaced by
$c/\sqrt{\varepsilon_{env}}$.

Finally the heat radiation of the object at temperature $T_{obj}$ can now be found by solving Eq.~\eqref{kruger2} for $C(T_{obj})$,
\begin{align}
&C(T_{obj})=a_{T_{obj}}(\omega) \frac{c^2}{\omega^2}\textrm{Im} \mathbb{G}-C^{env}(T_{obj})\label{eq:radfin},
\end{align}
where $\mathbb{G}$ is found using Eq.~\eqref{eq:GT}.
Note that $C^{env}(T)$ can be derived from either Eq.~\eqref{cenv} or directly from Eq.~\eqref{coldobject}. For the case of the cylinder, we present below the former derivation in detail, and briefly sketch the latter starting from Eq.~\eqref{coldobject}.

We emphasize again  that the field emitted by the object in Eq.~\eqref{eq:radfin} is fully expressed in terms of the Green's function with both arguments outside the object. In case one is only interested in the total heat emitted, the first term, i.e., the equilibrium field need not be derived, as it contains no Poynting vector. If, on the other hand, the radiation of the object is scattered at other objects, e.g. in order to compute heat transfer or nonequilibrium Casimir interactions, the full expression \eqref{eq:radfin} has to be kept.
\subsection{Heat radiation of a cylindrical object}\label{section2.2}
In order to compute the heat radiation of a  cylindrical object (denoted by subscripts $c$), we apply Eq.~\eqref{eq:radfin},  evaluating the environment contribution by use of Eq.~\eqref{cenv}. {Afterwards in this subsection}, we  briefly  sketch the derivation via Eq.~\eqref{coldobject}. In the cylindrical geometry (with parallel, radial and angular coordinates $z$, $r$ and $\phi$, respectively), the free Green's function $\mathbb{G}_0$ is expanded in cylindrical vector waves, $\textbf{RM}_{n,k_z}$ and $\textbf{RN}_{n,k_z}$ corresponding to $M$-polarized and $N$-polarized
regular waves \cite{Tsang}, see App. A. These are indexed by the multipole order $n$ and $k_z$, the component of the wavenumber $k=\omega/c$ along the cylindrical axis. For outgoing waves we use $\textbf{M}_{n,k_z}$ and $\textbf{N}_{n,k_z}$ accordingly. In this basis, the $\mathbb{T}$ operator of a cylindrical object is diagonal in $n$ and $k_z$, but couples different  polarizations. Its entry $T^{P'P}_{n,k_z}$ relates the amplitude of a scattered wave of polarization $P'$ in response to an incoming wave of unit amplitude and polarization $P$, with $P,P'\in \{M,N\}$. More precisely, the application of the $\mathbb{T}$ operator in Eq.~\eqref{eq:defT} on regular cylindrical functions reads,
%\begin{equation}\label{useful}
%\begin{split}
%&T^{PP'}_{n,k_z}2\pi \delta(k_z-k_z')\delta_{n,n'}=\frac{-i}{4}\int
%d^3r d^3r'\\
%&\textbf{RP}_{n,k_z}(r,-z,-\phi)\mathbb{T}_c(\textbf{r},\mathbf{r'})\textbf{RP}'_{n,k_z}(\textbf{r}').\\
%\end{split}
%\end{equation}
\begin{equation}
-\mathbb{G}_0\mathbb{T}_c \textbf{RP}_{n,k_z}=\sum_{P'} T^{P'P}_{n,k_z}\textbf{P}'_{n,k_z}.\label{eq:Tapp}
\end{equation}
With these definitions and Eq.~\eqref{eq:G0}, the Green's function of the cylinder is easily found, by use of Eq.~\eqref{eq:GT}, as
\begin{equation}\label{greenscyl}
\begin{split}
&\mathbb{G}_c=\mathbb{G}_0-\mathbb{G}_0\mathbb{T}_c\mathbb{G}_0=\mathbb{G}_0+\sum_{P,P'}\sum_{n=-\infty}^{\infty}(-1)^n\\
&\int_{-\infty}^{\infty}\frac{idk_z}{8\pi}\textbf{P}_{n,k_z}(\textbf{r})\otimes
\textbf{P}'_{-n,-k_z}(\textbf{r}')T^{PP'}_{n,k_z}.
\end{split}
\end{equation}
When performing the integration in Eq.~\eqref{cenv}, we note that $\mathbb{G}_0(\vct{r},\vct{r}')$ in Eq.~\eqref{eq:G0} is separated into two pieces, corresponding to $r<r'$ and $r'<r$.  {The contribution of a finite region vanishes asymptotically in the limit of $\varepsilon_{env}\to 1$ and can thus be neglected without changing the result}. We can hence restrict the integration range to $r'\geq r,r''$, where we have to use exclusively one of the pieces. In general, one can restrict the integration in Eq.~\eqref{cenv} to $\xi(\textbf{r}')\geq \xi(\textbf{r}),\xi(\textbf{r}'')$, where $\xi$ is the component which distinguishes the two expansions of $\mathbb{G}_0$.

 {Due to the orthogonality of two basis sets of the wave functions,} the integrations over polar angle $\phi'$ and cylindrical axis $z'$ yield $2\pi\delta_{n,n'}$ and $2\pi\delta(k_z-k'_z)$, respectively, and we are left with only one term for each polarization in Eq.~\eqref{cenv},
\begin{equation}
\begin{split}
\lim_{\varepsilon_{env}\to1}&\int
r'dr'\left|\mathbf{\widetilde{P}}_{-n,-k_z}\left(\textbf{r}'\right)\right|^2=\frac{2 c^2}{\pi \omega^2}\frac{1}{\textrm{Im}\varepsilon_{env}}+\dots,\
\end{split}
\end{equation}
where $\mathbf{\widetilde{P}}_{n,k_z}$  {has analogous form to} $\mathbf{P}_{n,k_z}$ with the  {wavenumber} $\sqrt{\varepsilon_{env}}\omega/c$ instead of $\omega/c$. Also, ``$\dots$'' represent higher order terms in $\varepsilon_{env}-1$. This equation holds for $k_z^2<\omega^2/c^2$, for $k_z^2>\omega^2/c^2$ all terms are of order
$\varepsilon_{env}^0$ and do not contribute in Eq.~\eqref{cenv},  a
manifestation of the fact that the environment radiation does not
contain evanescent waves.

The radiation from the environment after scattering at the cylinder then reads
\begin{widetext}
\begin{equation}\label{envrad}
\begin{split}
&C_c^{env}(T_{env})(\vct{r},\vct{r}'')=a_{T_{env}}(\omega)\sum_{P,P'}\sum_{n=-\infty}^{\infty}\int_{-\omega/c}^{\omega/c}\frac{dk_z}{8\pi}\frac{c^2}{\omega^2}\biggl\{\textbf{RP}_{n,k_z}(\textbf{r})
\otimes\textbf{RP}'^*_{n,k_z}(\textbf{r}'')\delta_{P,P'}+\textbf{RP}_{n,k_z}(\textbf{r})\otimes\textbf{P}'^*_{n,k_z}(\textbf{r}'')T_{n,k_z}^{P'P*}\\
&+\textbf{P}_{n,k_z}(\textbf{r})\otimes\textbf{RP}'^*_{n,k_z}(\textbf{r}'')T_{n,k_z}^{PP'}+\textbf{P}_{n,k_z}(\textbf{r})\otimes\textbf{P}'^*_{n,k_z}(\textbf{r}'')\sum_{P''}T_{n,k_z}^{PP''}T_{n,k_z}^{P'P''*}\biggl\},
%\left[\left(T_{n,k_z}^{PP}T_{n,k_z}^{P'P*}+T_{n,k_z}^{PP'}T_{n,k_z}^{P'P'*}\right)(1-\delta_{PP'})+
%\left(\left|T_{n,k_z}^{PP}\right|^2+\left|T_{n,k_z}^{P\overline{P}}\right|^2\right)\delta_{P,P'}\right]\biggl\},
\end{split}
\end{equation}
\end{widetext}
where $\overline{P}$ stands for the polarization opposite to $P$. Physically, Eq.~\eqref{envrad} describes the thermal field for the case of an environment at temperature $T_{env}$ and a cold cylinder ($T_c=0$). It can also be derived via Eq.~\eqref{coldobject}, by noting that the radiation of the environment without cylinder present can be given in closed form,
\begin{align}
\langle \vct{E}_0\otimes\vct{E}^*_0\rangle=a_{T_{env}}(\omega)\frac{c^2}{\omega^2}\Im\mathbb{G}_0=a_{T_{env}}(\omega)\sum_{P}\sum_{n=-\infty}^{\infty}\notag\\
\int_{-\omega/c}^{\omega/c}\frac{dk_z}{8\pi}\frac{c^2}{\omega^2}\left[\textbf{RP}_{n,k_z}\otimes\textbf{RP}^*_{n,k_z}\right].\label{eq:new}
\end{align}
Application of $(1-\mathbb{G}_0\mathbb{T}_c)$ from both sides (compare Eq.~\eqref{coldobject}), using Eq.~\eqref{eq:Tapp} immediately leads to Eq.~\eqref{envrad}.  {This simple route towards the environment radiation (and hence the radiation of the object) has not been presented before.}

Since we are only interested in the energy emitted by the cylinder, we do not explicitly derive the equilibrium field in Eq.~\eqref{eq:radfin} as it contains no Poynting vector. Equation \eqref{eq:radfin} thus states Kirchhoff's law, that the energy absorbed by the cylinder in the case $T_{env}=T$, $T_c=0$ is the same as the energy radiated by it for $T_{env}=0$, $T_c=T$. This is a special case of detailed balance in equilibrium, which generally states that the absorption coefficient of an object equals its emission coefficient. The formalism described here hence also provides a convenient route to find the absorption coefficient of arbitrary objects. Technically, due to these considerations, the Poynting vector
\begin{equation}\label{poynting2}
\langle\textbf{S}(\textbf{r})\rangle=\frac{c}{4\pi}\int\frac{d\omega}{(2\pi)^2}\Re\left[\langle\textbf{E}(\omega,\textbf{r})\times\textbf{H}^*(\omega,\textbf{r})\rangle\right],
\end{equation}
of the field in Eq.~\eqref{envrad} gives complete information about the net energy flux for any temperature combinations. It
can be derived via
$\textbf{B}(\omega,\textbf{r})=\frac{-ic}{\omega}\mathbf{\nabla}\times\textbf{E}(\omega,\textbf{r})$  as well as relation \eqref{useful2}.
The power $|H_c|$ radiated per
length $L$ of the infinite cylinder in the general case of finite $T_{env}$ and $T_c$ is finally given by~\cite{Kruger},
\begin{equation}\label{radiation}
\begin{split}
&\frac{|H_c|}L=-\frac{\hbar}{\pi^2}\int_{0}^{\infty}\omega d\omega\left[\frac 1{e^{\frac{\hbar\omega}{k_BT_c}}-1}-\frac 1{e^{\frac{\hbar\omega}{k_BT_{env}}}-1}\right]
\\&\sum_{P=M,N}\sum_{n=-\infty}^{\infty}
\int_{-\omega/c}^{\omega/c} dk_z (\mathrm
{Re}[T_{n,k_z}^{PP}]+|T_{n,k_z}^{PP}|^2+|T_{n,k_z}^{P\overline{P}}|^2).\
\end{split}
\end{equation}
Obviously, if $T_c<T_{env}$, there is a net energy flux into the cylinder. In the following, we consider exclusively the case $T_{env}=0$ (and denote $T_c=T$), for which case the energy flux is referred to as heat radiation of the cylinder.
The expression \eqref{radiation} can be split into two terms each representing a different polarization of the corresponding electric field. Specifically, the term which
describes polarization parallel to the cylinder is given by the $N$-modes,
\begin{equation}\label{radiationN}
\begin{split}
&\frac{|H_N|}L=\frac{|H_{\parallel}|}L=-\frac{\hbar}{\pi^2}\int_{0}^{\infty}\frac{\omega d\omega}{e^{\frac{\hbar\omega}{k_BT}}-1}\\
&\sum_{n=-\infty}^{\infty}\int_{-\omega/c}^{\omega/c} dk_z (\mathrm
{Re}[T_{n,k_z}^{NN}]+|T_{n,k_z}^{NN}|^2+|T_{n,k_z}^{NM}|^2),\
\end{split}
\end{equation}
whereas the term responsible for the polarization perpendicular to
the cylindrical axis is given by the $M$-modes,
\begin{equation}\label{radiationM}
\begin{split}
&\frac{|H_M|}L=\frac{|H_{\perp}|}L=-\frac{\hbar}{\pi^2}\int_{0}^{\infty}\frac{\omega d\omega}{e^{\frac{\hbar\omega}{k_BT}}-1}\\
&\sum_{n=-\infty}^{\infty}\int_{-\omega/c}^{\omega/c} dk_z (\mathrm
{Re}[T_{n,k_z}^{MM}]+|T_{n,k_z}^{MM}|^2+|T_{n,k_z}^{MN}|^2).\
\end{split}
\end{equation}
In the following we use a standard definition of the degree of polarization $I$ in order to quantify polarization effects,
\begin{equation}\label{polarization}
I=\frac{|H_N|-|H_M|}{|H_N|+|H_M|}.
\end{equation}
In case one prefers a description in terms of the scattering matrix $\mathbb{S}$ \cite{Rahi,bimonte1} with $\mathcal{S}_{n,k_z}^{P'P}=2T_{n,k_z}^{P'P}+ {\delta_{P,P'}}$, the radiation in Eq.~\eqref{radiation} can equivalently be written
\begin{equation}\label{Mohammad}
\begin{split}
\frac{|H_c|}L&=-\frac{\hbar}{4\pi^2}\int_{0}^{\infty}\omega d\omega\left[\frac 1{e^{\frac{\hbar\omega}{k_BT_c}}-1}-\frac 1{e^{\frac{\hbar\omega}{k_BT_{env}}}-1}\right]\\&\sum_{P,P'}\sum_{n=-\infty}^{\infty}
\int_{-\omega/c}^{\omega/c} dk_z (
|\mathcal{S}_{n,k_z}^{P P'}|^2- {\delta_{P,P'}}).\
\end{split}
\end{equation}
Furthermore we note that the result for the perpendicular emission of the cylinder given in Ref.~\cite{bimonte1} can be recovered from Eq.~\eqref{radiation} by restricting to $k_z=0$, taking into account waves normal to the cylindrical axis only. In this case the $\mathbb{T}$ operator is diagonal in polarization $P$.

\subsection{Limit of large radius (radiation of a plate of anisotropic material)}\label{section2.3}
For large radius, the radiation of a cylinder is asymptotically identical to the radiation of a plate (a semi-infinite planar object) of same surface area \cite{Kruger}. For the case of a plate made of isotropic material, the heat radiation is well-known \cite{Rytov}. Nevertheless, we will below study the heat radiation of a cylinder made of uniaxial material (as a simple model for carbon nanotubes), see Fig.~\ref{cylfig}, in which case the limit of large radius is a plate of uniaxial material. Recent literature discusses heat transfer between plates of uniaxial materials~\cite{uniaxialplates} as well as Casimir forces between a uniaxial plate and single-walled carbon nanotubes~\cite{swcntplate}, but we have not come across an explicit result for radiation of a plate. For materials with anisotropic electric or magnetic response, the Fresnel coefficients are not diagonal in polarization $s$ and $p$ (see Ref.~\cite{uniaxialcoef} for these lengthy coefficients), but take the general form $r^{Q'Q}$ for a scattered wave of polarization $Q'$ in response to an incoming wave of polarization $Q$. Note, $s(p)$ polarization corresponds to the wave with the electric field vector perpendicular (parallel) to the plane of incidence. Thus, the heat radiated per surface area (the Poynting vector $S$) can in this general case easily be found from Eq.~\eqref{eq:radfin},  {where, using a plane-waves basis \cite{Tsang}, the steps are similar to the ones performed for the cylinder (Eqs.~\eqref{greenscyl}-\eqref{envrad}).}
\begin{align}\label{plateresult}
S&=\frac {\hbar}{8\pi^3} \int_0^\infty d\omega \notag\frac{\omega}{e^{\frac{\hbar\omega}{k_BT}}-1}\\&\int_{k_\perp<\omega/c} d^2k_\perp \sum_{Q=\{s,p\}}\left[1-\left(|r^{Q}|^2+|r^{Q\bar{Q}}|^2\right)\right],
\end{align}
Here, $k_\perp$ is the wave-vector component parallel to the plate. For $r^{Q\bar{Q}}=0$, this equation reduces to the well-known one for isotropic materials \cite{Rytov}.
The expression \eqref{plateresult} can be rewritten in terms of $M$ and $N$ polarization for cylindrical geometry.
If we define $\phi$ to be the angle between the optical axis (which, in order to  {describe the radiation of a thick cylinder in terms of the one of a plate}, is parallel to the plate surface) and the intersection line between the plane of incidence and plate itself, then we can write the limiting values of the $M$ and $N$  {components of the cylinder radiation} as,
\begin{align}\label{SM}
&S_M=\frac {\hbar}{8\pi^3} \int_0^\infty d\omega \notag\frac{\omega}{e^{\frac{\hbar\omega}{k_BT}}-1}\int_0^{2\pi}d\phi\int_{0}^{\omega/c} k_\perp d k_\perp \\& \times\left[1-\frac{1}{2} \left\{\left(|r^{s}|^2+|r^{sp}|^2\right)\cos^2\phi+\left(|r^{p}|^2+|r^{ps}|^2\right)\sin^2\phi\right\}\right].
\end{align}
\begin{align}\label{SN}
&S_N=\frac {\hbar}{8\pi^3} \int_0^\infty d\omega \notag\frac{\omega}{e^{\frac{\hbar\omega}{k_BT}}-1}\int_0^{2\pi}d\phi\int_{0}^{\omega/c} k_\perp d k_\perp\\&\times \left[1-\frac{1}{2} \left\{\left(|r^{s}|^2+|r^{sp}|^2\right)\sin^2\phi+\left(|r^{p}|^2+|r^{ps}|^2\right)\cos^2\phi\right\}\right].
\end{align}
Note $r^{sp}=r^{ps}$ when the optical axis is parallel to the plate.
\section{$\mathbb{T}$ operator for  a cylinder made of uniaxial material}\label{section3}
\begin{figure}\centering
\subfigure{\label{cylfig}}
\includegraphics[width=8.5 cm]{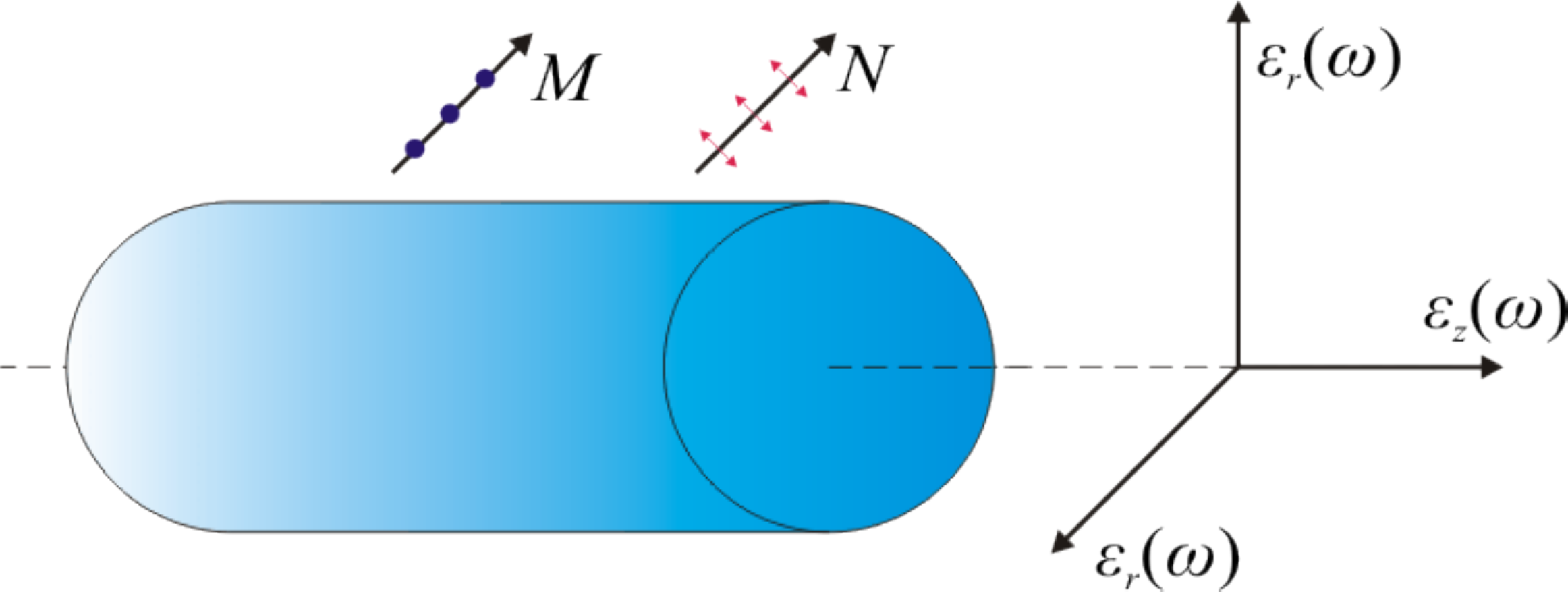}
\vspace*{-0.5cm}
\caption{(color online) Infinitely long
cylinder made of a uniaxial material.
The
symmetry axis of the cylinder coincides with the symmetry axis of the diagonal dielectric tensor (the $z$-axis).
Electromagnetic waves radiated by the cylinder are denoted by $M$
and $N$ for $M$-polarized (perpendicular) and
$N$-polarized (parallel) respectively. Note, generally $M$-polarized waves have components along both azimuthal and radial directions. Also, despite $N$-polarized waves have non-zero components along all three basis directions, it is only the component parallel to the cylindrical axis to contribute to the Poynting vector.}\label{cylfig}
\end{figure}
In section~\ref{section2.2}, we derived the heat radiation of a cylindrical object expressed in terms of its $\mathbb{T}$ operator which is known analytically \cite{Bohren,thorsten}.
One aim of this paper is to study the radiation of a cylinder of uniaxial material as a simple model for carbon nanotubes (see Sec. \ref{section4.3}). The corresponding $\mathbb{T}$ operator, which seems unavailable in the literature, will be derived here for the case depicted in Fig.\ref{cylfig}, where the cylindrical axis coincides with the optical axis. This is
done by solving the scattering problem in Eq.~\eqref{eq:defT}, which amounts to satisfying the boundary conditions for the electromagnetic field at the
cylinder's surface. The setup is described by isotropic local magnetic permeability $\mu(\omega)$ and the following tensor for the anisotropic, but homogeneous and local dielectric function inside the cylinder,
\begin{equation}\label{tensor}
\widehat\varepsilon(\omega)=\left( \begin{array}{ccc}
\varepsilon_{r}{(\omega)} & 0 & 0 \\
0 & \varepsilon_{r}{(\omega)} & 0 \\
0 & 0 & \varepsilon_z{(\omega)} \end{array} \right).
\end{equation}
The electric displacement field inside the cylinder can then be written as
$$\textbf{D}=\varepsilon_{r}{(\omega)}\left(E_r\mathbf{e_{r}}+E_{\phi}\mathbf{e_\phi}\right)+\varepsilon_{z}(\omega)E_z\mathbf{e_z},$$
where $\mathbf{e}_r$, $\mathbf{e}_\phi$ and $\mathbf{e}_z$
correspond to unit vectors in cylindrical coordinates.

Importantly, uniaxial materials split the incident beam into
ordinary and extraordinary ones~\cite{Born}. In our geometry, cylindrical $M$-polarized waves correspond to ordinary ones and propagate according to the dielectric function $\varepsilon_r$. The $N$ modes correspond to extraordinary waves and propagate according to an effective dielectric function which depends on the direction of propagation.

The resulting $\mathbb{T}$ operator components $T_{n,k_z}^{P'P}$, as defined in Eq.~\eqref{eq:Tapp}, are
expressed in terms of Bessel functions, $J_n$, and Hankel functions
of first kind, $H^{(1)}_n$ (see App.~\ref{B} for the detailed derivation),
\begin{equation}\label{TMMA}
T_{n,k_z}^{MM} =-\frac{J_n
(qR)}{H_n^{(1)}(qR)}\frac{\Delta_1\Delta_4-K^2}{\Delta_1\Delta_2-K^2},\\
\end{equation}
\begin{equation}\label{TEEA}
T_{n,k_z}^{NN}=-\frac{J_n
(qR)}{H_n^{(1)}(qR)}\frac{\Delta_2\Delta_3-K^2}{\Delta_1\Delta_2-K^2},\\
\end{equation}
\begin{equation}\label{TMEA}
T_{n,k_z}^{NM}=T_{n,k_z}^{MN}=\frac{2i}{\pi\sqrt{\varepsilon_z\mu}(qR)^2}\frac{K}{[H_n^{(1)}(qR)]^2}\frac{1}{\Delta_1\Delta_2-K^2},
\end{equation}
%\begin{equation}\label{TEMA}
%T_{n,k_z}^{MN}=\frac{2i}{\sqrt{\varepsilon_z}\pi(qR)^2}\frac{K}{[H_n^{(1)}(qR)]^2}\frac{1}{\Delta_1\Delta_2-K^2},
%\end{equation}
where
\begin{equation}\label{delta1}
\Delta_1=\frac{J'_n(q_NR)}{q_NRJ_n(q_NR)}-\frac 1 {\varepsilon_{z}}
\frac{H_n^{(1)'}(qR)}{qRH_n^{(1)}(qR)},
\end{equation}
\begin{equation}\label{delta2}
\Delta_2=\frac{J'_n(q_MR)}{q_MRJ_n(q_MR)}-
\frac 1 {\mu}\frac{H_n^{(1)'}(qR)}{qRH_n^{(1)}(qR)},
\end{equation}
\begin{equation}\label{delta3}
\Delta_3=\frac{J'_n(q_NR)}{q_NRJ_n(q_NR)}-\frac 1 {\varepsilon_z}
\frac{J'_n(qR)}{qRJ_n(qR)},
\end{equation}
\begin{equation}\label{delta4}
\Delta_4=\frac{J'_n(q_MR)}{q_MRJ_n(q_MR)}-
\frac 1 {\mu}\frac{J'_n(qR)}{qRJ_n(qR)},
\end{equation}
and
\begin{equation}\label{delta5}
\begin{split}
%&K_{1}=K_{NM}=\frac{nk_zc}{\sqrt{\varepsilon_r}R^2\omega}\left[\left(\frac
%1 {q_M^2}-\frac 1 {q^2}\right)\left(\frac 1 {q_{N}^2}-\frac \alpha { q^2}\right)\right]^{1/2}.\\
&K=\frac{nk_zc}{\sqrt{\varepsilon_z\mu} R^2\omega}\left(\frac
1 {q_M^2}-\frac 1 {q^2}\right).\\
\end{split}
\end{equation}
$k=\omega/c$ and
$q=\sqrt{k^2-k_z^2}$ are the wave-vector magnitude in vacuum and its
component perpendicular to the $z$-axis respectively.
$q_M=\sqrt{\varepsilon_{r}\mu k^2-k_z^2}$ and
$q_N=\sqrt{\varepsilon_{z}/\varepsilon_{r}}\sqrt{\varepsilon_{r}\mu
k^2-k_z^2}$ are the wave-vector components perpendicular to the
$z$-axis for the $M$-polarized ordinary and $N$-polarized
extraordinary waves inside the cylinder respectively.
%$\widetilde{k}_z=k_z/k$.

As required by continuity, the above $\mathbb{T}$ matrix  can be easily reduced to the
isotropic case when $\varepsilon_{r}=\varepsilon_z=\varepsilon$. Then the expressions
simplify to $q_M=q_N=\sqrt{\varepsilon\mu
k^2-k_z^2}\equiv q'$ and
$K=nk_zc(1/q'^2-1/q^2)/(\sqrt{\varepsilon\mu}R^2\omega)$, and our results reduce to the known forms for an isotropic cylinder \cite{Bohren,thorsten}.

\section{Examples and asymptotic results}\label{section4}

In this section we numerically, and analytically, study the radiation of a cylinder for different material classes. We start with isotropic dielectrics and
conductors and finally present the case of a uniaxial material using the ``in-layer'' and ``inter-layer'' response of graphite to model MWCNT. We consider $\mu=1$ for all studied materials.
\subsection{Dielectric cylinder}\label{section4.1}
%\subsubsection{SiO$_2$ and SiC cylinders}
\begin{figure}\centering
\subfigure[]{
\includegraphics[width=7 cm]{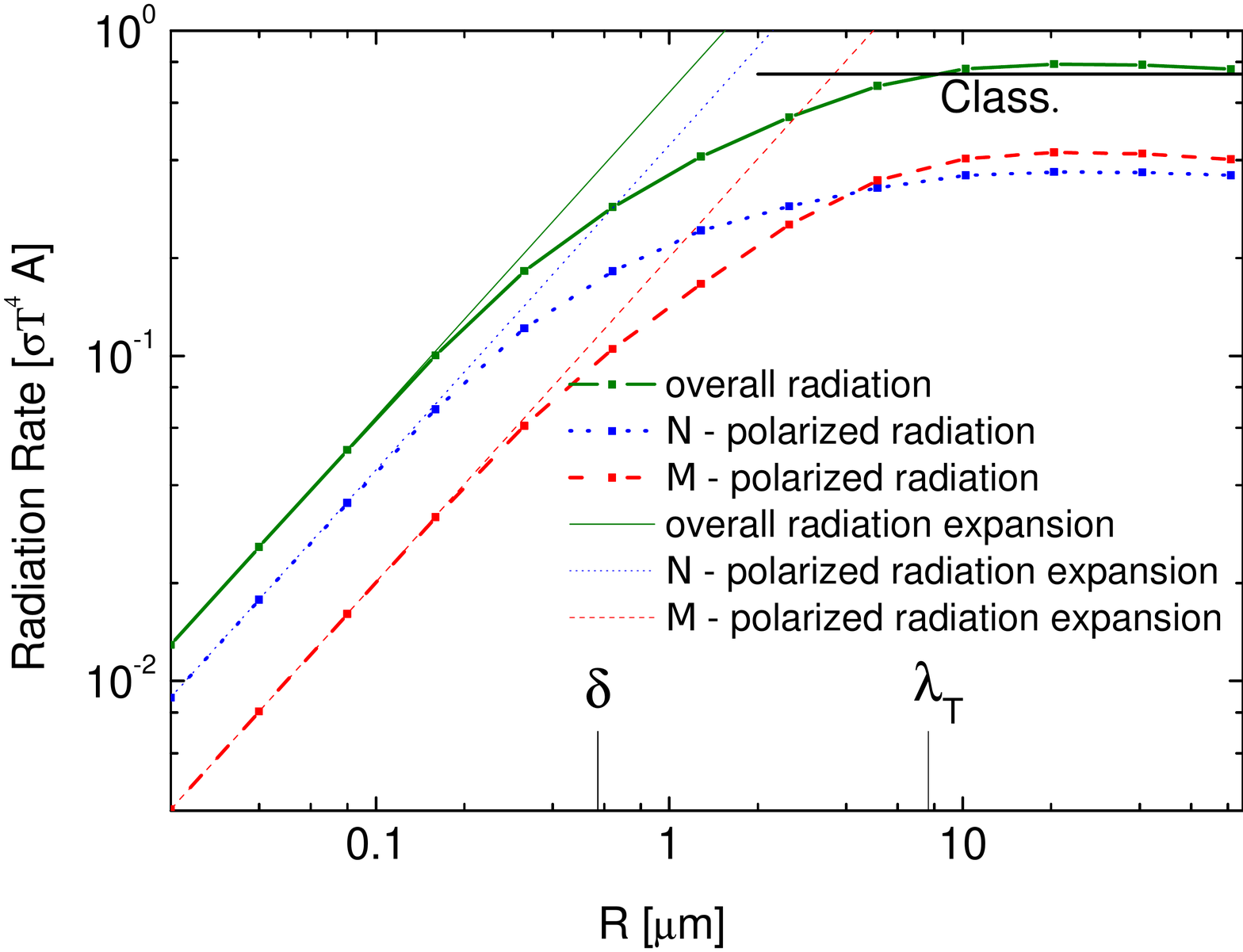}
\vspace*{-0.5cm}
\label{SiO2} } \subfigure[]{
\includegraphics[width=7 cm]{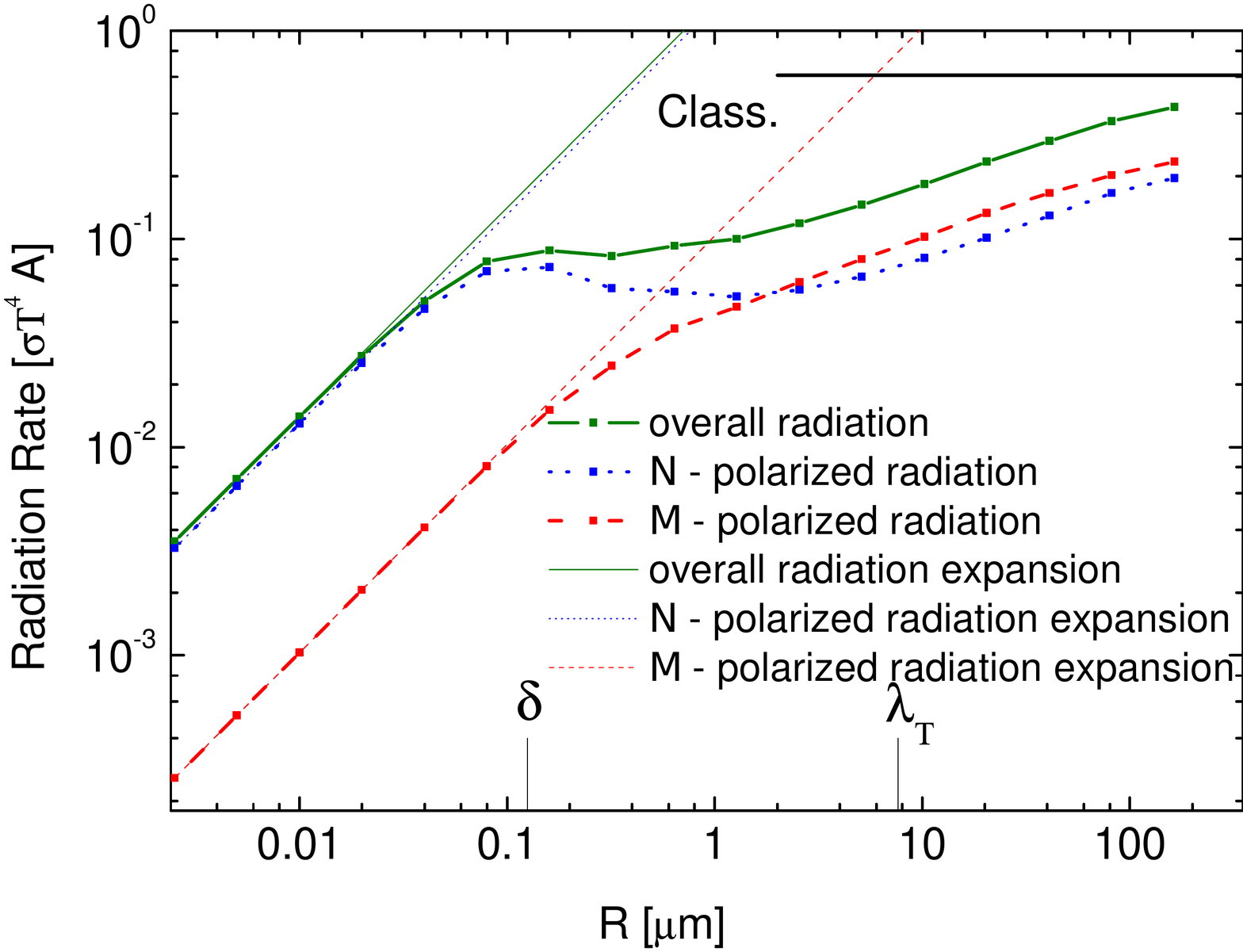}
\vspace*{-0.5cm}
\label{SiC} } \caption{(color online) The normalized heat radiation as a function of
radius $R$ for (a) SiO$_2$ and (b) SiC cylinders at
temperature $T=300K$. Calculations were performed using
Eq.~\eqref{radiation}, and analytical expansions,
Eq.~\eqref{cylErad} and Eq.~\eqref{cylMrad}. The horizontal lines
show the radiation of (a) SiO$_2$ and (b) SiC plates. $\lambda_T$
and the smallest skin depths $\delta$ in the relevant frequency
range are marked on $R$-axis.}\label{diel}
\end{figure}

Figure \ref{diel}
%~\ref{SiO2}%
illustrates the result of the numerical calculation of the radiation by SiO$_2$ and SiC cylinders for $T=300K$ normalized by the Stefan-Boltzmann value, Eq.~\eqref{SB}.

For SiO$_2$ optical data was used, whereas for SiC the following dielectric function was taken~\cite{ChenNew},
\begin{equation*}
\varepsilon_{SiC}(\omega)=\varepsilon_\infty\frac{\omega^2-\omega_{LO}^2+i\omega\gamma}{\omega^2-\omega_{TO}^2+i\omega\gamma},
\end{equation*}
where $\varepsilon_\infty=6.7$, $\omega_{LO}=0.12eV$, $\omega_{TO}=0.098eV$, $\gamma=5.88\times10^{-4}eV$.

We see the effects discussed in Ref.~\cite{Kruger} involving the three length scales: radius $R$, thermal wavelength $\lambda_T$, and skin depth $\delta=c/(\Im\sqrt{\varepsilon}\omega)$, where the latter depends on frequency. If $R$ is the smallest scale, i.e., much smaller than the smallest \emph{relevant skin depth} and $\lambda_T$, the radiation is proportional to the volume of the cylinder. In this case, radiation emitted inside the cylinder will hardly be reabsorbed on its way out so that all regions of the cylinder contribute equally to the emission.  The other asymptotic behavior is approached when $R$ is the largest scale, i.e. much larger than the largest \emph{relevant skin depth} and $\lambda_T$. Then only the cylinder surface contributes to the radiation which approaches the values of a plate of equal surface area as seen in the figure.

The length scale $\lambda_T$ sets (via the Boltzmann factor) the range of relevant wavelengths of emission. Nevertheless, for dielectrics there is a fine structure to this range given by the resonances of the material. In case of $R\ll\delta$, the cylinder emits predominantly at the resonance wavelengths of the material, where $\delta$ is minimal. On the other hand, for $R\gg \{\delta,\lambda_T\}$, the emissivity is strongest in regions where $\varepsilon\approx 1$ (compare the plate emissivity).

In general, one might expect resonance effects when the emitted wavelength is of the order of $R/(2\pi)$ (similar to Mie resonances for a sphere \cite{Bohren, Kattawar70}). Due to the contribution of all wavelengths, these are smeared out in the total heat emitted. For SiO$_2$ in Fig.~\ref{diel}, the fact that the emissivity exceeds the plate result for $R\approx20\mu$m might be connected to such resonances.

Figure \ref{diel} shows imprints of the dielectric function of SiC which has a sharp strong resonance (leading to a small $\delta$), but apart from the resonance, SiC is almost black (i.e. has very large $\delta$) in our frequency range. The two regimes $R\ll\{\delta,\lambda_T\}$ and $R\gg\{\delta,\lambda_T\}$, where the cylinder radiation follows the discussed asymptotic laws are far apart, and very large radii $R$ are necessary to approach the classical plate result. In Fig.~\ref{diel}, the radiation for $R=150\mu$m is still distinctly different from the asymptotic result. This might be advantageous for experiments, as a SiC cylinder does not have to be very thin in order to observe deviations from the Stefan-Boltzmann law.

In the limit of $R\ll \{\delta,\lambda_T\}$, the radiation can be studied analytically, see App.~\ref{app:an}, where we find the following asymptotic laws for the two polarizations,
\begin{equation}\label{cylErad}
\begin{split}
\lim_{R\ll\{\delta,\lambda_T\}}\frac{|H_N|}A&=\frac 1 6 \int_{0}^{\infty}\frac{d\omega}{(2\pi)^4}a_T
(\omega) c
R\\ &\times\textrm{Im}\frac{\varepsilon(\omega)^2+2\varepsilon(\omega)-1}{\varepsilon(\omega)+1},\\
\end{split}
\end{equation}
\begin{equation}\label{cylMrad}
\begin{split}
\lim_{R\ll\{\delta,\lambda_T\}}\frac{|H_M|}A=\frac 1 2\int_{0}^{\infty}\frac{d\omega}{(2\pi)^4}a_T
(\omega) c
R\textrm{Im}\frac{\varepsilon(\omega)-1}{\varepsilon(\omega)+1}.
\end{split}
\end{equation}
Note the linear increase with $R$, corresponding to the proportionality of the unnormalized radiation to the volume.
The numerical evaluation of these asymptotic forms has been added to the graphs in Fig.~\ref{diel}, where the agreement for small $R$ to the full results is visible. As expected and seen from the equations above,
cylinders with purely real dielectric functions (or more general with real potential $\mathbb{V}$) will not radiate, because the dissipative properties of the material are responsible
for the heat radiation (in accord with FDT). This holds for any $R$. The $N$-polarization given in Eq.~\eqref{cylErad} dominates over the $M$-polarization in  Eq.~\eqref{cylMrad} if
\begin{equation}\label{TEElimit2}
[\Re\varepsilon(\omega)+1]^2+[\Im\varepsilon(\omega)]^2\gg
\Im\varepsilon(\omega)
\end{equation}
holds, which is the case for most materials. Further insight can be gained by additionally requiring the temperature to be so low that one can expand the dielectric function, i.e., $\lambda_T\gg\lambda_0$, where $\lambda_0$ is the
wavelength of the lowest resonance of the material. In this
case~\cite{Jackson},
\begin{equation}\label{EpsilonLowT}
\varepsilon(\omega)=\varepsilon_0+i\frac{\lambda_{in}\omega}{c}+\mathcal{O}(\omega^2),
\end{equation}
with  $\varepsilon_0$ and $\lambda_{in}$ real. Plugging
\eqref{EpsilonLowT} into Eqs.\eqref{cylErad} and \eqref{cylMrad} the frequency integration can be done and we have,
\begin{equation}\label{cylEradLowT}
\lim_{R\ll\{\delta,\lambda_T\},\lambda_0\ll\lambda_T}\frac{|H_N|}A=\frac{4\pi^4}{189}\frac{\hbar
c^2\lambda_{in}R}{\lambda_T^6}\left[1+\frac{2}{(\varepsilon_0+1)^2}\right],
\end{equation}
\begin{equation}\label{cylMradLowT}
\lim_{R\ll\{\delta,\lambda_T\},\lambda_0\ll\lambda_T}\frac{|H_M|}A=\frac{8\pi^4}{63}\frac{\hbar
c^2\lambda_{in}R}{\lambda_T^6}\frac{1}{(\varepsilon_0+1)^2}.
\end{equation}
Interestingly, to lowest order in $T$, both components scale as $T^6$ and hence fundamentally different from the Stefan-Boltzmann law in Eq.~\eqref{SB}, which scales as $T^4$.

The degree of polarization $I$ in
Eq.~\eqref{polarization} is  finally given by
\begin{equation}\label{degreepol}
\lim_{R\ll\{\delta,\lambda_T\},\lambda_0\ll\lambda_T}I=\frac{\varepsilon_0^2+2\varepsilon_0-3}{\varepsilon_0^2+2\varepsilon_0+9}.
\end{equation}
Although the condition $\lambda_0\ll\lambda_T$ is not fulfilled in Fig.~\ref{diel}, Eq.~\eqref{degreepol} still gives a hint to why SiC has higher degree of polarization compared to SiO$_2$, as $\varepsilon_0$ is tangibly larger for SiC and  Eq.~\eqref{degreepol} monotonically increases with $\varepsilon_0$ (for $\varepsilon_0>0$).
\subsection{Well conducting cylinder}\label{section4.2}

\begin{figure}\centering
\subfigure[]{
\includegraphics[width=7.0 cm]{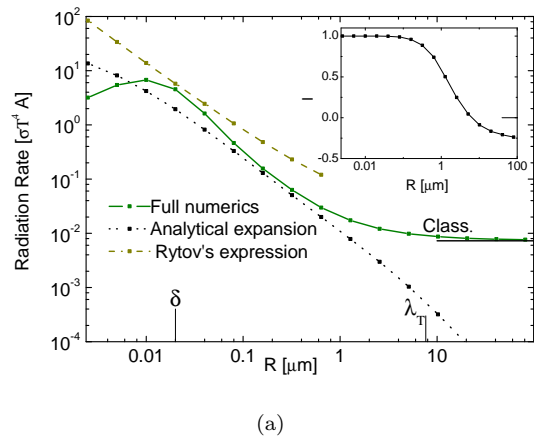}
\label{Gold300}} \subfigure[]{
\includegraphics[width=7.0 cm]{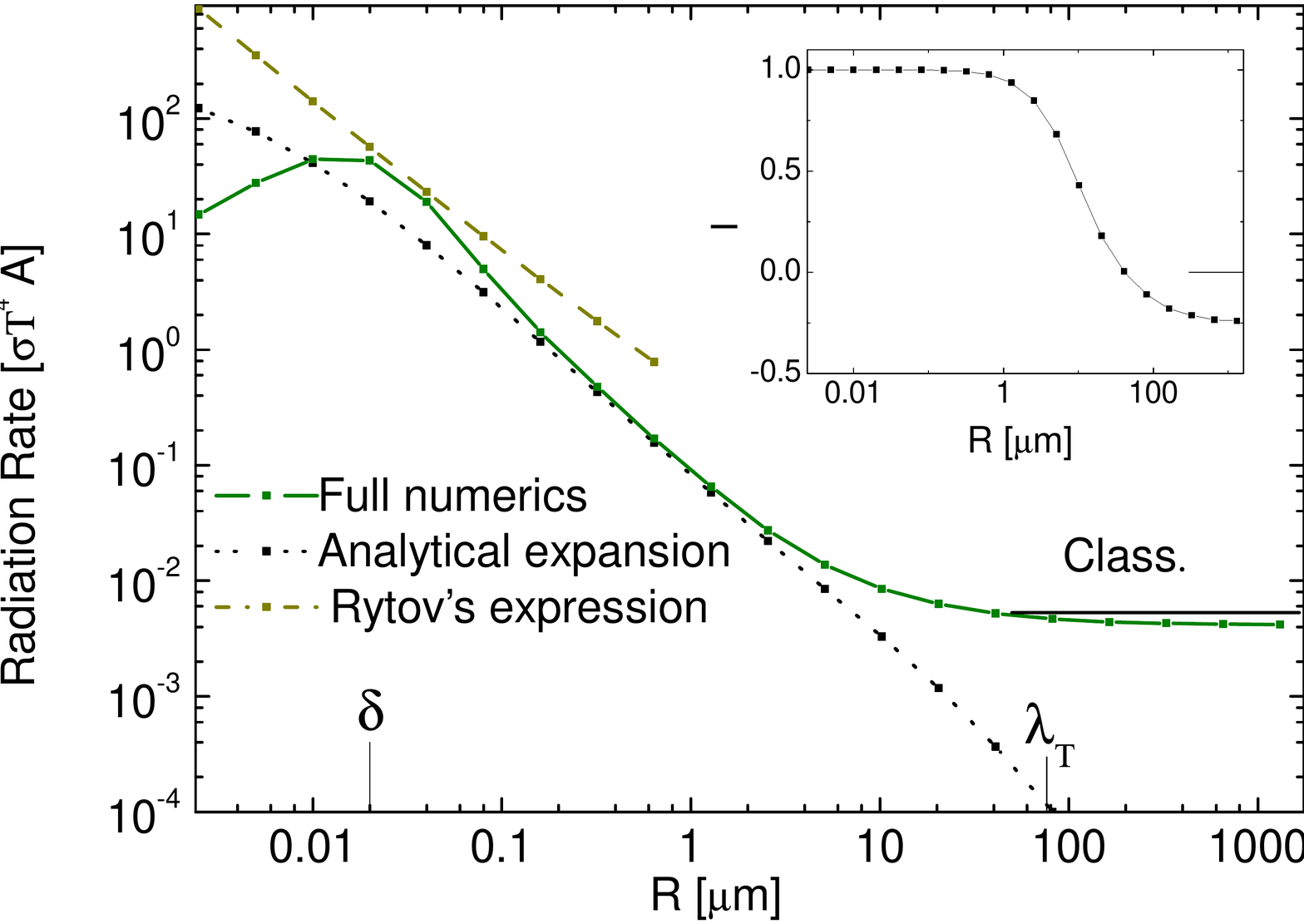}
\label{Gold30}} \caption{(color online) The normalized heat radiation as a function
of radius $R$ for Au cylinder at temperatures (a) $T=300K$
and (b) $T=30K$. Calculations were performed using the Drude
model~\cite{Gold} with Au optical properties. Full numerics,
Eq.~\eqref{radiation}, analytical expansion, Eq.~\eqref{condrad},
and the approximation of Eq.~\eqref{spectralfinalnormalized}
were used. Black horizontal lines indicate the radiation of Au
plates at corresponding temperatures. $\lambda_T$ and the skin depth
$\delta$ in the relevant frequency range are marked on the $R$-axis. In
the insets the degree of polarizations, \eqref{polarization}, are plotted.
}\label{Au}
\end{figure}

\begin{figure}\centering
\subfigure[]{
\includegraphics[width=7.0 cm]{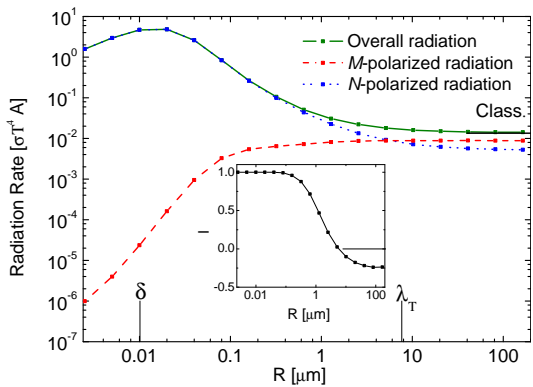}
\label{Tungsten298}} \subfigure[]{
\includegraphics[width=7.0 cm]{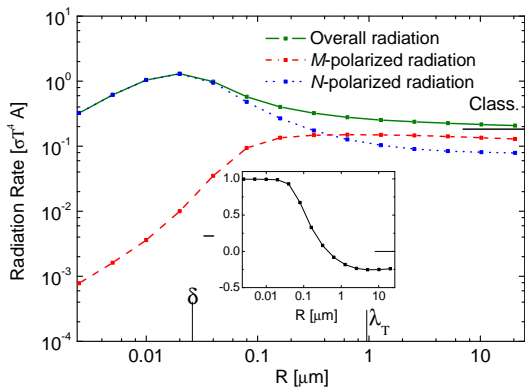}
\label{Tungsten2400}} \caption{(color online) The normalized heat radiation as a function
of radius $R$ for tungsten cylinder at temperatures (a) $T=298K$
and (b) $T=2400K$. Calculations were performed using Eq.\eqref{radiation} with dielectric function \eqref{Tdiel}. Black horizontal lines indicate the radiation of tungsten
plates at corresponding temperatures. $\lambda_T$ and the skin depth
$\delta$ in the relevant frequency range are marked on $R$-axis. In
the insets the  degree of polarizations are given using
expression \eqref{polarization}.}\label{Tungsten}
\end{figure}

Conductors differ from insulators by a significantly smaller
skin depth $\delta$, leading to very different radiation
characteristics. In this section, we first study the radiation using a simple Drude model with the parameters of gold in order to highlight the different limiting behaviors. Then, we turn to tungsten which has been extensively used in experiments\cite{Ohman,bimonte1}.

The Drude model for gold\cite{Gold},
\begin{equation}\label{drudegold}
\varepsilon_{Au}(\omega)=\varepsilon_\infty-\frac{\omega_p^2}{\omega(\omega+i\omega_\tau)},
\end{equation}
has the parameters $\varepsilon_\infty=1$, $\omega_p=9.03eV$ and $\omega_\tau=2.67\times10^{-2}eV$.

Figure~\ref{Au} shows the numerical result for the total radiation by a gold
cylinder at $300K$ and $30K$.  We observe a behavior drastically different from the ones in Fig.~\ref{diel}: the radiation, normalized by surface area, can be many orders larger than expected from the Stefan-Boltzmann law. More precisely, it increases with decreasing $R$, has a maximum at  $R\approx\delta$ (for Drude model of gold \eqref{drudegold}, $\delta(\omega)$ has no sharp minimum, and we show the skin depth corresponding roughly to thermal wavelength) and approaches the laws \eqref{cylErad} and \eqref{cylMrad} only for very small (unphysical) radii. For $R\gg\lambda_T$, the result of a gold plate is approached. The large values of radiated power, compared to the Stefan-Boltzmann law, can be
explained by the interplay of two effects: a large imaginary part of the dielectric
function gives rise to strong radiation from every volume element of the conductor. On the other hand, it also leads to a very small skin depth such that most radiation is reabsorbed inside the conductor, and only a thin surface layer contributes to the radiation. In the region where $R\approx \delta$, one has very strong emitting elements which all contribute to the total radiation, and hence emission normalized by surface area is maximal. Interestingly, when the conductivity goes to infinity, the effect of vanishing skin depth is stronger than the effect of increasing radiation such that the  emissivity vanishes as $1/\sqrt{\varepsilon}$. In this case, the classical (plate) limit goes to zero and the maximum in the curve shifts to smaller and smaller $R$.

The insets show the
degree of polarization~\eqref{polarization} as a function of $R$. For $R\alt\lambda_T$, the radiation from
the gold cylinder is fully $N$-polarized. For $R\agt \lambda_T$, the degree of polarization becomes negative and asymptotically approaches zero for $R/\lambda_T\to\infty$. These qualitative features agree with experimental\cite{Ohman,Agdur,Ingvarsson,Skulason,bimonte1} as well as theoretical\cite{bimonte1} studies on radiation of thin wires.

For conductors, the appropriate limit for an  analytic expansion is $\lambda_T\gg R\gg \delta$ (where $\delta$ is of the order of nanometers for good conductors). In this case, the leading element of the $\mathbb{T}$ operator is $T^{NN}_{0,k_z}$, see App.~\ref{D}. The resulting radiation is then completely $N$-polarized and reads
\begin{widetext}
\begin{equation}\label{condrad}
\lim_{\lambda_T\gg R\gg
\delta}\frac{|H_c|}A=\int_{0}^{\infty}\frac{d\omega}{(2\pi)^2}a_T (\omega)
\frac{c^2}{(2\pi)^2\omega} \int_{0}^{\pi/2} d\theta
\frac{2\mathrm{Re}[1/\sqrt{\varepsilon}]\cos^3{\theta}}{\left|\cos^2{\theta}(2\gamma_E-i\pi)\omega
R/c-2i/\sqrt{\varepsilon}+2\cos^2{\theta}\omega
R/c\log{[\cos{\theta}\omega R/2c]}\right|^2},
\end{equation}
\end{widetext}
where $\gamma_E\approx0.577$ is Euler-Mascheroni constant and $\theta$ is the angle of incidence so that $k_z=k\sin\theta$.

As described above, we see here explicitly that the radiation (in the considered range of radii) vanishes for $|\varepsilon|\to\infty$.
This expression cannot be further simplified as the integrand
diverges at $\theta=\pi/2$ if we omit the small term
$-2i/\sqrt{\varepsilon}$ in the denominator. A result almost similar to
\eqref{condrad} was obtained by Rytov~\cite{RytovCyl} (whose derivation was
restricted to $\textrm{Re}\varepsilon=0$ and
$\textrm{Im}\varepsilon\gg0$), which differs by the absence of the
first term in denominator. We emphasize however that this term is necessary for
accurate results in the considered limit.

Rytov suggests a further simplification of his expression by setting $\cos{\theta}=1$ in $\log{[\cos{\theta}\omega R/2c]}$,
so that integration can be performed analytically. Omitting $\cos^2{\theta}(2\gamma_E-i\pi)\omega
R/c$ in denominator, the following result for radiation can be obtained after integration,	
\begin{equation}\label{spectralfinalnormalized}
\begin{split}
\lim_{\lambda_T\gg R\gg
\delta}\frac{|H_c|}A&\approx
\int_0^{\infty}\frac{d\omega}{(2\pi)^2}\frac{c^{7/2}}{2(2\pi)^2\omega^{5/2}}\\
&\times a_T (\omega)\mathrm{Re}[1/\sqrt{\varepsilon}]
\frac{|\varepsilon|^{1/4}}{|R\log{[\omega R/2c]}|^{3/2}}.\
\end{split}
\end{equation}
Note that Eq.~\eqref{spectralfinalnormalized} was derived for any
well-conducting (non magnetic) media, whereas the corresponding result, Eq. (IV.35) in Ref.~\cite{RytovCyl}, only considers the case of
a purely imaginary dielectric function (and $\mu\not=1$). Due to this the integrand of Eq.~\eqref{spectralfinalnormalized} differs from Eq. (IV.35) in Ref.~\cite{RytovCyl} by $\mathrm{Re}[1/\sqrt{\varepsilon}]
|\varepsilon|^{1/4}$ instead of $(\mu/4 \Im\varepsilon)^{1/4}$ (which agree for $\Re\varepsilon=0$ and $\mu=1$). Also, we emphasize that  Eq.~\eqref{spectralfinalnormalized} is an approximation whereas Eq.~\eqref{condrad} is exact asymptotic limit for $\lambda_T\gg R\gg \delta$.

Figure \ref{Au} provides a test of these approximations, demonstrating that for $T=30K$, where the ratio of $\lambda_T$ and $\delta$ is larger than at $T=300K$, Eq.~\eqref{condrad} describes the full solution over roughly one decade in $R$. For $T=300K$, the agreement is not as good as the ratio of $\lambda_T$ and $\delta$ is too small.
Equation \eqref{spectralfinalnormalized} gives a rough estimate of the overall dependence on $R$ in Fig.~\ref{Au}, but the values differ by roughly a factor of 10 from the exact results. Moreover,  {above} some threshold value of  $R$ we cannot
obtain a finite radiation rate from Eq.~\eqref{spectralfinalnormalized}
because of the divergent term due to log of unity in the denominator
of the integrand.

After this (more theoretical) discussion of gold, where we used the same $\varepsilon$ for both temperatures in order to demonstrate the pure effect of temperature via the Boltzmann factor, we turn to the experimentally relevant material tungsten, at relevant temperatures of $T=298K$ and $T=2400K$ as shown in Fig.~\ref{Tungsten}. We use the corresponding dielectric function for tungsten\cite{Wdiel},
\begin{equation}\label{Tdiel}
\varepsilon_W(\omega)=1+\sum_{p=1}^3\frac{K_{0p}\lambda^2}{\lambda^2-\lambda_{sp}^2+i\delta_p\lambda_{sp}\lambda}-\frac{\lambda^2}{2\pi c\epsilon_0}\sum_{q=1}^2\frac{
\sigma_q}{\lambda_{rq}-i\lambda},
\end{equation}
where $\lambda$ is the wavelength in vacuum, $c$ is the velocity of light and $\epsilon_0$ is the permittivity of vacuum in SI units.
The remaining parameters are listed in Table \ref{table1}.
\begin{table}
\begin{tabular}{|c|c|c|}
  \hline
  $T$ & 298 & 2400 \\
  \hline$\sigma_1$ & 17.5 & 1.19 \\
 \hline $\sigma_2$ & 0.21 & 0.25 \\
 \hline $\lambda_{r1}$ & 45.5 & 3.66 \\
 \hline $\lambda_{r2}$ & 3.7 & 0.36 \\
 \hline $K_{01}$ & 12 &  \\
 \hline $K_{02}$ & 14.4 &  \\
 \hline $K_{03}$ & 12.9 &  \\
  \hline$\lambda_{s1}$ & 1.26 &  \\
 \hline $\lambda_{s2}$ & 0.6 &  \\
 \hline $\lambda_{s3}$ & 0.3 &  \\
  \hline$\delta_1$ & 0.6 &   \\
 \hline $\delta_2$ & 0.8 &   \\
 \hline $\delta_3$ & 0.6 &   \\
  \hline
\end{tabular}
\caption{Optical data for tungsten from \cite{Wdiel}. Temperature $T$ is in Kelvins. Conductivities ($\sigma_1$, etc.)
are in units of $10^6$ ohm$^{-1}$m$^{-1}$. Wavelengths ($\lambda_{r1}$, etc.) are in
$\mu$m. $K_{01}$, etc. and $\delta_1$, etc. are dimensionless.}\label{table1}
\end{table}

The overall radiation of tungsten at $298K$ is very similar to gold at $300 K$ in Fig.~\ref{Au}. At high temperature $T=2400 K$, the increase of the normalized radiation over the Stefan-Boltzmann law is reduced. We attribute this to a smaller conductivity at $T=2400 K$. We also observe in the insets that the zero in the polarization curves shifts by roughly a factor of 10 when comparing the two temperatures. This manifests that the zero in the polarization curves is mostly a function of $R/\lambda_T$ as $\lambda_T$ is also shifted by almost a factor of 10. Furthermore, although the polarization in the inset is indistinguishable from unity at small $R$, we note that the ratio of $N$ and $M$ polarizations is a factor of $10^3$ larger for $T=298 K$ compared to $T=2400 K$.

\subsection{Multi-walled carbon nanotube}\label{section4.3}
\begin{figure}\centering
\subfigure{\label{graphite}}
\includegraphics[width=8.5 cm]{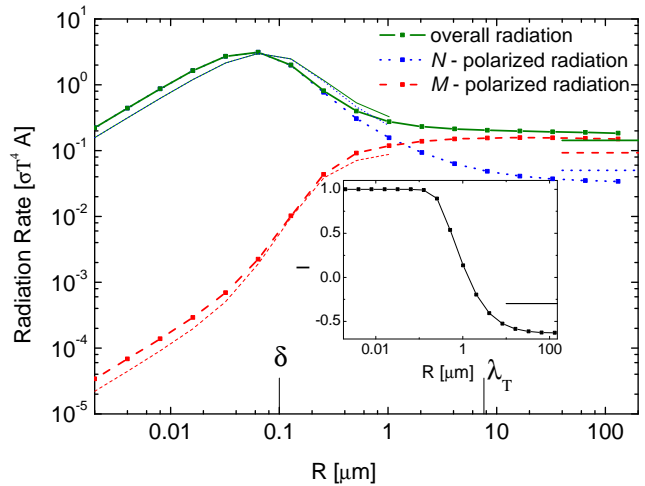}
%\vspace*{-0.5cm}
\caption{(color online) The heat radiation of a MWCNT as a
function of radius $R$, normalized by the Stefan-Boltzmann result,
at $T=300 K$. Contributions from two polarizations are indicated. Corresponding thin curves without boxes represent the heat radiation for ``isotropic graphite'' with dielectric function $(\varepsilon_z+\varepsilon_r)/2$.
Horizontal lines of different colors indicate the graphite plate classical result for correspondingly colored curves.
$\lambda_T$ and the smallest skin depth $\delta$ in the relevant
frequency range are marked on the $R$-axis. Note that the smallest
skin depths corresponding to both $\varepsilon_r$ and
$\varepsilon_z$ are approximately equal and labeled here by
$\delta$. In the inset the degree of polarization is given
using  expression \eqref{polarization}.}\label{graphite}
\end{figure}
We finally turn to heat radiation of a cylinder made of uniaxial material. This can be considered a simple model for a MWCNT\cite{prikladuha} which is of high
importance in modern science and technology. As a MWCNT is a wrapped up graphite sheet, we can in a crude approximation regard it as a (solid) cylinder described by two different dielectric response functions: the response along the cylindrical axis is given by the in-layer properties of the graphite sheets, whereas the response perpendicular to this axis is approximately given by the inter-layer response. We also note that most mineral crystals have uniaxial optical properties, and heat emission by these materials might open new possibilities for applications\cite{uniaxialplates}.

Figure \ref{graphite} shows the heat radiation of a MWCNT for $T=300K$ using expressions
\eqref{radiation} and \eqref{TMMA}-\eqref{delta5}, normalized as before by the Stefan-Boltzmann law $H=\sigma T^4 A$. We used the following form for the graphite dielectric function\cite{serbka},
\begin{equation}\label{dielgraphite}
\varepsilon_{r,z}(\omega)=1-\frac{\Omega_p^2}{\omega\left(\omega+i\Gamma_0\right)}-\sum_j\frac{f_j\omega_p^2}{(\omega^2-\omega_{tj}^2)+i\omega\Gamma_j'},
\end{equation}
where $\Omega_p=\sqrt{f_0}\omega_p$ and a specific form of $\Gamma_j'=\Gamma_j\exp\left[-\alpha_j\left(\frac{\hbar\omega-\hbar\omega_j}{\Gamma_j}\right)^2\right]$ is used, which best describes the experimental data. For the dielectric function component $\varepsilon_z$ along the cylindrical axis, i.e., the in-layer response, the parameters are $\omega_p=19 eV$, $\Gamma_0=0.091 eV$, $f_0=0.016$. The remaining parameters are given in Table \ref{table2}. For the dielectric function component $\varepsilon_r$ perpendicular to the cylindrical axis, i.e., the inter-layer response, the parameters are $\omega_p=27 eV$, $\Gamma_0=6.365 eV$, $f_0=0.014$ and the remaining parameters are given in Table \ref{table3}.
\begin{table}
\begin{tabular}{|c|c|c|c|c|c|c|c|}
  \hline
   $j$ & 1 & 2 &  3 & 4 & 5 &   6 &  7 \\
 \hline
  $f_j$ &0.134 & 0.072&  0.307& 0.380& 0.065& 0.553& 1.381 \\
 \hline
  $\alpha_j$ &24.708& 0.524& 0.217& 0.518& 0.286& 0.248& 15.101 \\
\hline
  $\omega_{tj}$ &2.358& 5.149& 13.785& 10.947& 16.988& 24.038& 36.252 \\
\hline
$\Gamma_j$ & 9.806 & 472.7 & 4.651 & 1.797 & 2.418 & 21.395 & 37.025\\
\hline
\end{tabular}
\caption{Optical parameters for the in-layer  dielectric function of  graphite ($\varepsilon_z$) from \cite{serbka}. $f_j$ and $\alpha_j$ are dimensionless, whereas $\omega_{tj}$ and $\Gamma_j$ are in $eV$.}\label{table2}
\end{table}

\begin{table}
\begin{tabular}{|c|c|c|c|c|c|c|c|}
  \hline
   $j$ & 1 & 2 &  3 & 4 & 5 &   6 &  7 \\
 \hline
  $f_j$ &0.073& 0.056& 0.069& 0.005& 0.262& 0.460& 0.200\\
 \hline
  $\alpha_j$ &0.505& 7.079& 0.362& 7.426& 0.000382& 1.387& 28.963 \\
\hline
  $\omega_{tj}$ &0.275& 3.508& 4.451& 13.591& 14.226& 15.550& 32.011 \\
\hline
$\Gamma_j$ &4.102& 7.328& 1.414& 0.046& 1.862& 11.922& 39.091\\
\hline
\end{tabular}
\caption{Optical parameters for the inter-layer dielectric function of graphite ($\varepsilon_r$)  from \cite{serbka}. $f_j$ and $\alpha_j$ are dimensionless, whereas $\omega_{tj}$ and $\Gamma_j$ are in $eV$.}\label{table3}
\end{table}

These parameterizations apply to the frequency range $0.12-40eV$ %(\approx1.82\times10^{14}-6.07\times10^{16}Hz)$
and $2-40eV$ %(\approx3.03\times10^{15}-6.07\times10^{16}Hz)$
for $\varepsilon_r$ and $\varepsilon_z$ respectively, but we nevertheless use them for the range of roughly $0.004-0.2eV$%(\approx6.28\times10^{12}-3.14\times10^{14}Hz)$
due to the lack of simple analytic forms for the broader range.  Note that $\lambda_T$ corresponds to a frequency of $0.163eV$ for $T=300K$.

The overall radiation curve is in between the characteristic shapes of dielectrics and conductors (compare Figs~\ref{diel} and \ref{Au}): the regime proportional to volume as in Eqs.~\eqref{cylErad} and \eqref{cylMrad} is visible for small $R$ in contrast to Fig.~\ref{Au}. On the other hand, the strong increase over the plate-result and over the Stefan-Boltzmann value, characteristic for conductors, is visible as well. These features follow from the dielectric functions in Eq.\eqref{dielgraphite} carrying  a smaller conductivity compared to gold.

In order to highlight the effect of material anisotropy, in Fig.~\ref{graphite} we also show the thin curves without boxes for which we use an isotropic dielectric function, given by $(\varepsilon_z+\varepsilon_r)/2$. We see that the influence of anisotropy on the radiation is almost negligible at small $R$, whereas at intermediate $R$ it strongly increases the degree of polarization perpendicular to the tube. Interestingly, the asymptotic value of $I$ in the inset is different from 0 and takes the value $-0.297$, an effect purely due to the anisotropy which is computed using the result for a plate of anisotropic material, Eqs.\eqref{SM} and \eqref{SN}. We note that a very thick MWCNT might in fact be best described by a plate with optical axis \emph{perpendicular} to the surface. Thus, while Fig.~\ref{graphite} gives the correct radiation for a material with the dielectric properties given in Eq.~\eqref{dielgraphite}, it probably only describes MWCNT for small $R$. Since, at small $R$, the polarization is hardly dependent on the anisotropy of the material, we conclude that the polarization effects for MWCNT \cite{Li} are mainly an effect of cylindrical geometry rather than anisotropy of the material.

\section{Spectral emissivity}\label{section5}
In Sec.~\ref{section4}, we studied the total heat radiation of a cylinder made of different materials. While this is of interest in connection with efficient heating or cooling, another quantity which can be more appropriate for direct comparison to experiments is the spectral emissivity. In this section, we discuss  the spectral emissivity for cylinders made of dielectrics (SiO$_2$), conductors (tungsten) and uniaxial materials (graphite) for a fixed radius of $R=5\mu$m.

The spectral  {emissivity (density)} $H_\omega$ is given by the integrand
of Eq.~\eqref{radiation},
\begin{equation}\label{RadiationDensity}
\begin{split}
&\frac{|H_\omega|}L=-a_T (\omega)
\frac{c^4}{(2\pi)^4\omega^3}\sum_{P=M,N}\sum_{n=-\infty}^{\infty}\\
&\int_{-\omega/c}^{\omega/c} dk_z (\mathrm
{Re}[T_{n,k_z}^{PP}]+|T_{n,k_z}^{PP}|^2+|T_{n,k_z}^{P\overline{P}}|^2).\
\end{split}
\end{equation}
We denote by $H_{\omega M}$ and $H_{\omega N}$ the correspondingly polarized components of $H_\omega$.

\subsection{SiO2 }\label{section5.1}
\begin{figure}\centering
\includegraphics[width=7.0 cm]{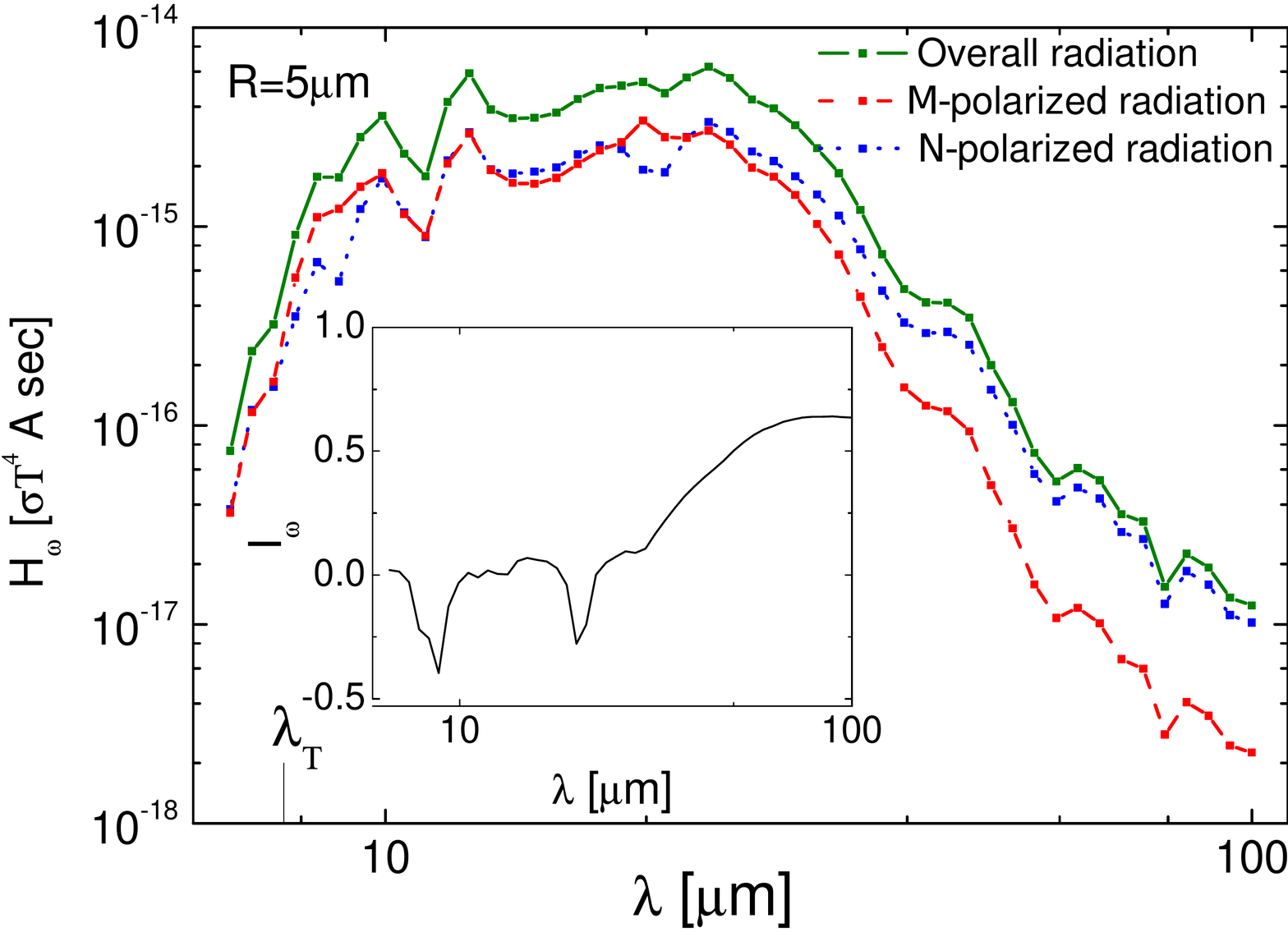}
\caption{(color online) The spectral density divided by Stefan-Boltzmann law as a function of
wavelength $\lambda$ for SiO$_2$ cylinder of radius $R=5\mu$m
at temperature $T=300K$. In the inset the spectral degree of
polarization is provided.}\label{SpecSiO2}
\end{figure}

Figure~\ref{SpecSiO2} illustrates the spectral density for
SiO$_2$ at $T=300K$. The unsteady local fine structure reflects the nature of the optical data which has a number of smaller peaks. For short wavelengths (high
frequencies) $M$-polarized radiation mostly dominates, whereas for
$\lambda\agt25 \mu$m the $N$-polarized
radiation starts to prevail up to the limit of long wavelengths (low frequencies). In the inset the spectral
degree of polarization as a function of wavelength is shown, where
\begin{equation}\label{specpol}
I_\omega=\frac{|H_{\omega N}|-|H_{\omega M}|}{|H_{\omega N}|+|H_{\omega M}|}.
\end{equation}
The two large valleys in this curve are due to the two dominant resonances of SiO$_2$. The fact that the resonances lead to negative valleys in the polarization is a feature specific for the radius chosen.
For large wavelengths, the spectral degree of
polarization approaches a constant value, which can be
computed easily using expressions \eqref{cylErad} and
\eqref{cylMrad},
\begin{equation}\label{pollimit}
\lim_{\{\delta,R\}\ll\lambda}I_\omega=\frac{|\varepsilon(\omega)|^2+2\Re\varepsilon(\omega)-3}{|\varepsilon(\omega)|^2+2\Re\varepsilon(\omega)+9}.
\end{equation}
Furthermore, if additionally $\lambda\gg\lambda_0$ holds, the dielectric function is described by Eq.\eqref{EpsilonLowT} and the spectral degree of polarization is given then by Eq.\eqref{degreepol}.

Note that the spectral degree of polarization is independent of
temperature if the dielectric function is independent of temperature.

\subsection{Tungsten}\label{section5.2}

\begin{figure}\centering
\subfigure[]{
\includegraphics[width=7.0 cm]{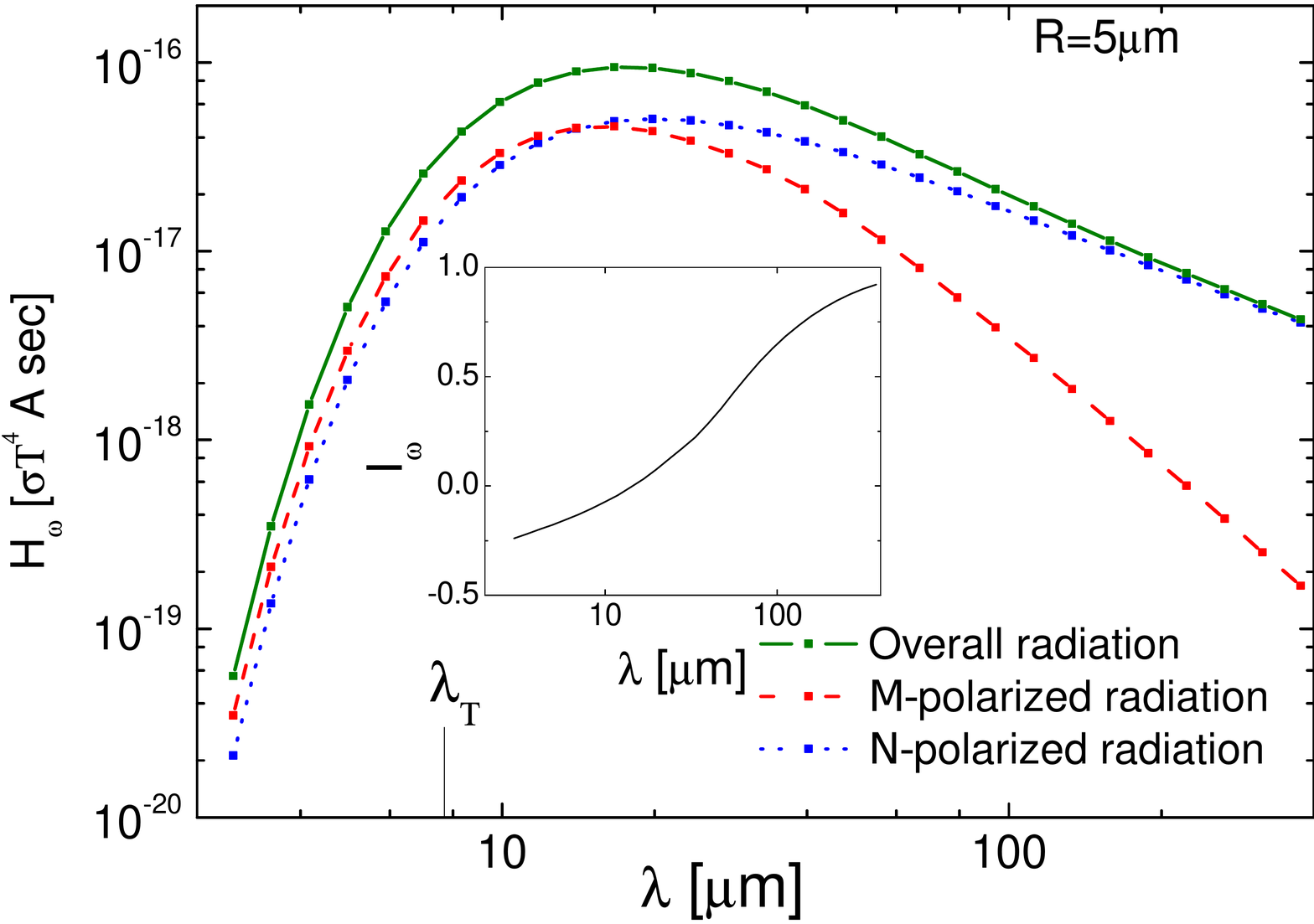}
\label{SpecTungsten298}} \subfigure[]{
\includegraphics[width=7.0 cm]{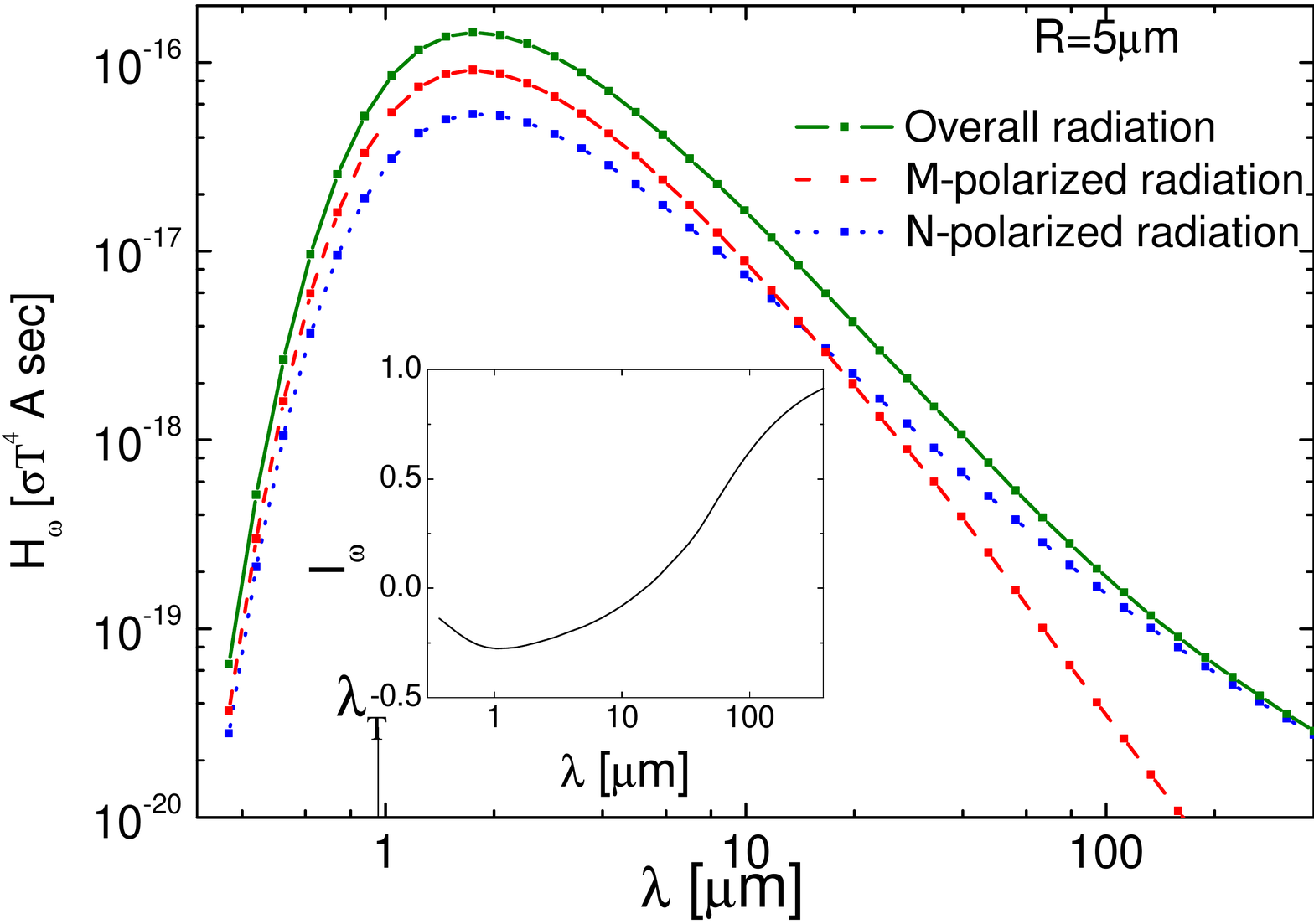}
\label{SpecTungsten2400}} \caption{(color online) The spectral
density divided by Stefan-Boltzmann law as a function of wavelength $\lambda$ for tungsten cylinder of
radius $R=5\mu$m at temperatures (a) $T=298K$ and (b)
$T=2400K$. In the insets the spectral degrees of polarization are
provided.}\label{SpecTungsten}
\end{figure}

Figure \ref{SpecTungsten} shows the spectral density of radiation
for  tungsten cylinders at $T=298K$ and $T=2400K$. The
shape of the curves is very similar to Planck's classical law due to the
Bose-Einstein statistics factor in $a_T(\omega)$. The curves peak at values slightly larger than the
corresponding thermal wavelengths. For short wavelengths
$M$-polarized radiation is stronger than $N$-polarized one,
whereas for $\lambda\agt 2R$, $N$-polarized radiation dominates, strongly suppressing the $M$-polarized radiation in the limit of long
wavelengths.
This transition of polarization  is
also manifested in the insets where spectral degrees of polarizations are plotted. In the
limit of long wavelengths these  approach
unity, as can be justified by results of Sec.\ref{section4.2},
where it is shown that the $N$ component is dominant in
the limit $\lambda_T\gg R\gg\delta$. On the other hand, in the limit of short wavelengths the spectral degree of polarization approaches zero as the cylinder radiates as an isotropic plate. We emphasize again that the spectral degree of polarization
is independent of temperature (if $\varepsilon(\omega)$ is).

\subsection{Graphite}\label{section5.3}
\begin{figure}\centering
\includegraphics[width=7.0 cm]{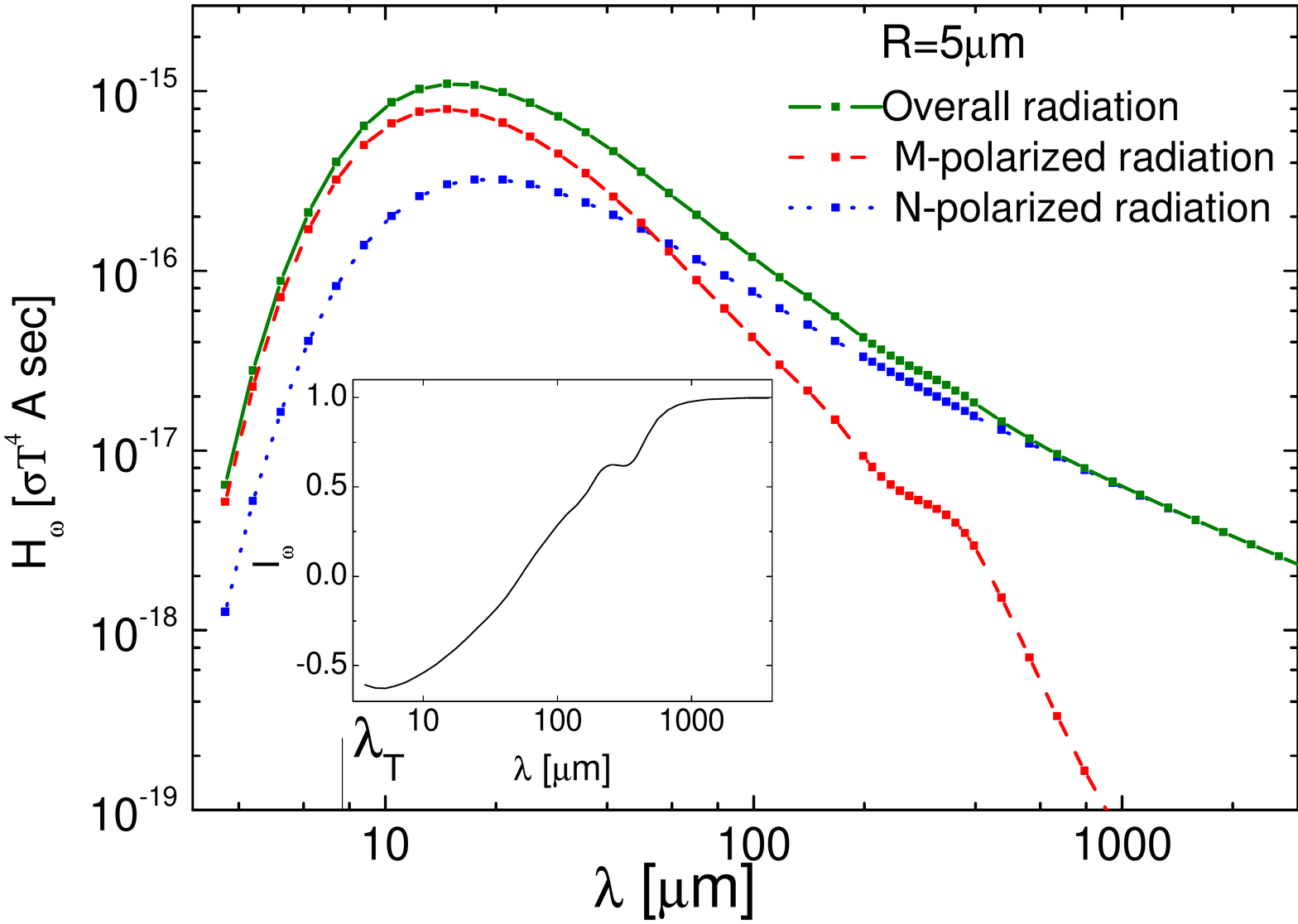}
\caption{(color online) The spectral density divided by Stefan-Boltzmann law as a function of
wavelength $\lambda$ for graphite cylinder of radius $R=5\mu$m
at temperature $T=300K$. In the inset the spectral degree of
polarization is provided.}\label{SpecGraphite300}
\end{figure}

Figure~\ref{SpecGraphite300} shows corresponding results for a graphite cylinder at $T=300K$, displaying similar behavior as the case of conductors in Fig.\ref{SpecTungsten}. The transition point, where  the polarization changes sign is at $\lambda\approx25 \mu$m. Analogously to conductors, the spectral degree of
polarization for a graphite cylinder tends to unity in the limit of
long wavelengths, i.e.  spectral density has polarization
parallel to the cylinder. The range of wavelengths $200-300\mu$m shows an unexpected wiggle in the curves, the origin of which is unclear.

We note that $I_\omega$ goes to zero for $\lambda\to0$, although a finite value is approached for $R/\lambda_T\to\infty$ in Fig.~\ref{graphite}. This is due to the fact that both $\varepsilon_z$ and $\varepsilon_r$ tend to 1 for $\omega\to\infty$, and the material is asymptotically isotropic. Nevertheless, at $\lambda\approx5\mu$m, the polarization is very strong compared to Fig.~\ref{SpecTungsten}, an effect which we attribute to the anisotropy of the material.

Another model describing the spectral degree of polarization of MWCNT's is presented in Ref.~\cite{regan2}, where we note partly common structure to our description arising from the expansions of Eqs.~\eqref{TdielectricTEE0}-\eqref{TdielectricTME1}. For large $\lambda/R\to \infty$, both the experimental data as well as the theoretical predictions of Ref.~\cite{regan2} give values for $I_\omega$ close to unity, in agreement to Fig.~\ref{SpecGraphite300}.

\subsection{Comparing material classes}\label{section5.4}
\begin{figure}\centering
\includegraphics[width=7.0 cm]{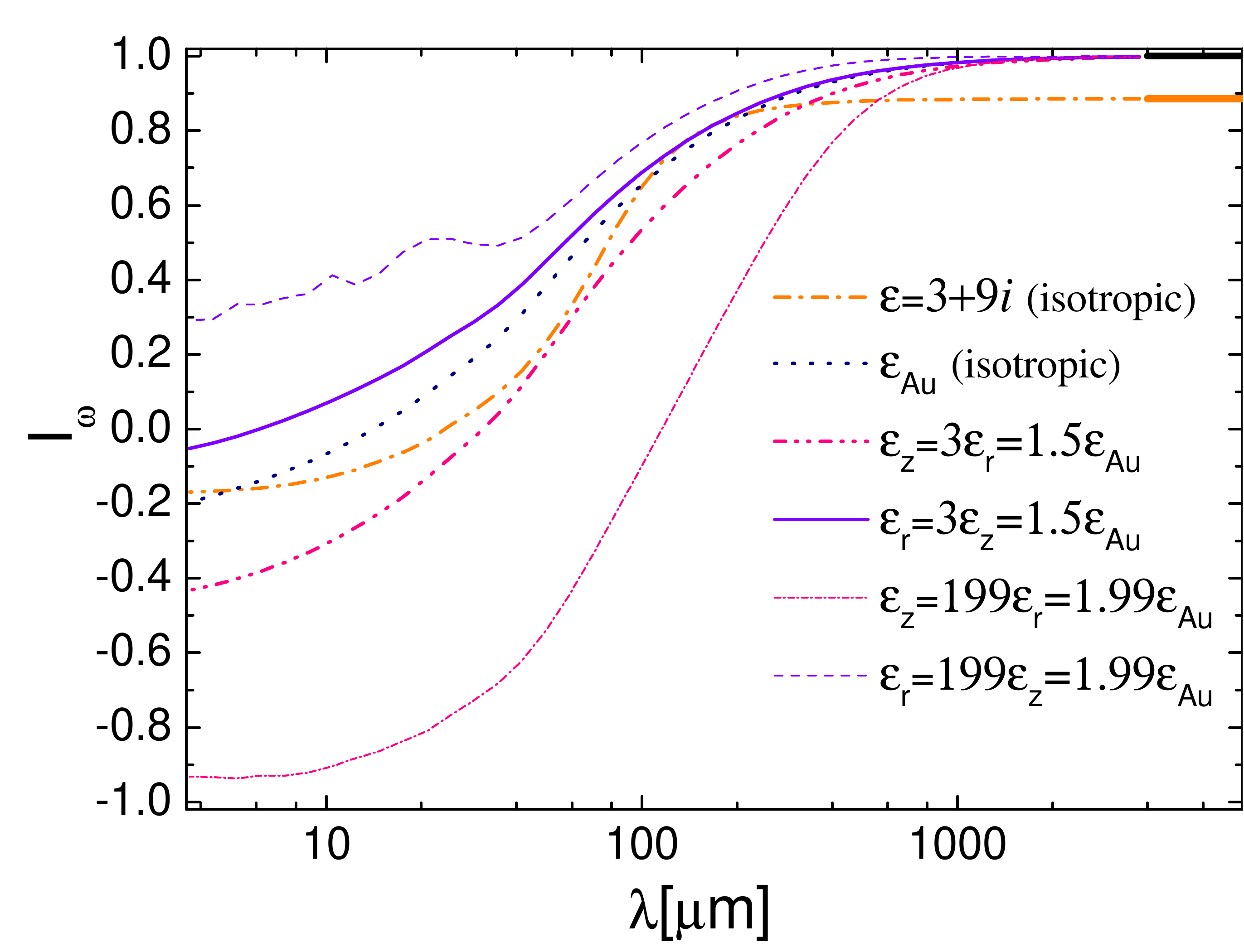}
\caption{(color online) Spectral degree of polarization as a function of
wavelength $\lambda$ for cylinders made of isotropic dielectric, isotropic conducting, and
 anisotropic conducting materials for  $R=5\mu$m. Black horizontal and orange lines correspond to
long wavelengths limiting values for conductors (isotropic and uniaxial) and the dielectric (using Eq.~\eqref{pollimit}), respectively.}\label{polset}
\end{figure}

Finally,  Fig.~\ref{polset} compares the spectral polarization for the different materials discussed, where we used simplified dielectric functions for dielectrics and uniaxial materials in order to illuminate the pure influence of $\lambda/R$.

Figure~\ref{polset} manifests that the overall dependence of the polarization on wavelength is quite universal following similar curves for all cases shown. Fundamental differences are seen in the limiting cases. In the limit of large wavelengths all conductors approach unity, whereas dielectrics go to a constant value different from unity when $\lambda\gg\{\delta,R\}$ \big{(}see Eq.~\eqref{pollimit}\big{)}.

In the opposite limit of small wavelengths, Fig.~\ref{polset} manifests strong dependence of the polarization on uniaxiality. We emphasize that by showing additionally artificial materials with very strong anisotropy (factor 199 between $\varepsilon_r$ and $\varepsilon_z$). Varying  this factor, a range from roughly 0.4 to -0.9 in polarization at $\lambda=10\mu$m can be sweeped. Figure~\ref{polset} clearly illustrates that the smaller $\varepsilon_r/\varepsilon_z$, the smaller the spectral polarization.

\section{Discussion}\label{section6}

In this article,  {we derived the general formalism to study} the heat radiation of a single object held at uniform temperature. While the presented  formalism can describe the radiation of arbitrary objects, we focus on cylindrical objects, providing explicitly the heat radiation formula (Eq. \eqref{radiation}) in terms of its $\mathbb{T}$ operator. In order to  {work out} the radiation of multi-walled carbon nanotubes in a simple model, we  {derive} the $\mathbb{T}$ operator for a cylinder made of uniaxial material. We study the cases of dielectrics, conductors and MWCNT numerically, discussing certain limits (e.g. small radius) analytically. To lowest order in temperature $T$, the heat emitted by dielectric cylinders is proportional to $T^6$ in contrast to $T^4$ in the Stefan-Boltzmann law, Eq.~\eqref{SB}.

For all materials, the limiting value of very large radius is given by the radiation of a plate of equal material. In the limit of small radius, dielectrics emit proportional to their volume, whereas this regime is not reached for conductors in the physically relevant  range of radii. Instead, the radiation normalized per surface area can be very large for conductors and we observe values as high as almost a hundred times the value expected from the Stefan-Boltzmann law. The maximum occurs where the radius is roughly equal to the skin depth (e.g. a few nm for gold). We note that the validity of using a continuous dielectric function at the scale of nanometers is questionable and has to be addressed in future studies.

The heat radiated by a long cylinder is polarized. All studied materials show common overall features of their degree of polarization: cylinders with radius smaller than the thermal wavelength emit predominantly radiation polarized parallel to their axis, whereas for radii larger than the thermal wavelength, the polarization points into the direction perpendicular to the cylinder. The limiting value of the degree of polarization for small radii is different for insulators and conductors: while the former approach a value less than unity, thin conductors emit completely polarized radiation and the degree of polarization approaches unity.

The effect of uniaxiality on the polarization is most dominant for large radii, where the degree of polarization can be tuned over a wide range by adjusting the ratio of parallel and perpendicular responses. For small radii, uniaxiality hardly influences the polarization. As MWCNT fall into the latter size regime, we conclude that the polarization measured experimentally for nanotubes are mostly a consequence of geometry, and not so much of material anisotropy.

Detailed comparison to experimental data (see e.g. Refs.~\cite{regan1,regan2,regan3} for MWCNTs) will be left for future work.

\begin{acknowledgments}
We thank T.~Emig, R.~L.~Jaffe, G.~Bimonte, M.~F.~Maghrebi, M.~T.~H.~Reid and N.~Graham for helpful discussions. We are especially grateful to T.~Emig for providing us with unpublished notes on $\mathbb{T}$ matrix for an isotropic cylinder and to M.~F.~Maghrebi for pointing us to the possibility of writing the radiation in the form of Eq.~\eqref{Mohammad}. This research was supported by the NSF Grant No.  DMR-08-03315, DARPA contract No. S-000354 and the DFG grant No. KR 3844/1-1.
\end{acknowledgments}
\numberwithin{equation}{section}
\appendix
\section{Cylindrical harmonics and Free Green's function in cylindrical
basis}\label{A}

According to Ref.~\cite{Tsang}, the cylindrical
harmonics can be written as,
$$\textbf{RM}_{n,k_z}(\textbf{r})=\left[\frac{in}{qr}J_n(qr)\mathbf{e}_r-J'_n(qr)\mathbf{e}_\phi\right]e^{ik_zz+in\phi},$$
\begin{equation}\label{cylindricalharmonics}
\begin{split}
\textbf{RN}_{n,k_z}(\textbf{r})&=\frac c{\omega}\left[ik_zJ'_n(qr)\mathbf{e}_r-\frac{nk_z}{qr}J_n(qr)\mathbf{e}_\phi\right.\\&\left.+qJ_n(qr)\mathbf{e}_z\right]e^{ik_zz+in\phi},\\
\end{split}
\end{equation}
where $J_n$ is the Bessel function of order $n$.
$\textbf{RM}_{n,k_z}$ and $\textbf{RN}_{n,k_z}$ correspond
to regular magnetic multipole (TE) and electric multipole (TM) waves
respectively. Also, $k_z$ and $q$ are the wavevectors parallel and
perpendicular to the cylindrical $z$-axis respectively satisfying
the relation $q=\sqrt{k^2-k_z^2}$, $ k=\omega/c$.
$J'_n$ corresponds to the first derivative with respect to the
argument. Furthermore, we denote the corresponding outgoing waves by
$\textbf{M}_{n,k_z}$ and $\textbf{N}_{n,k_z}$, which differ from
regular ones by replacing $J_n$ with the Hankel function of the
first kind $H_n^{(1)}$.

The above solutions correspond to transverse waves, i.e.
$\mathbf{\nabla}\cdot\mathbf{RM}_{n,k_z}=\mathbf{\nabla}\cdot\mathbf{RN}_{n,k_z}=0$.
Moreover, they obey useful relations
$\mathbf{RM}_{n,k_z}=\frac{c}{\omega}\mathbf{\nabla}\times\mathbf{RN}_{n,k_z}$,
$\mathbf{RN}_{n,k_z}=\frac{c}{\omega}\mathbf{\nabla}\times\mathbf{RM}_{n,k_z}$.
These relations are also valid for outgoing waves.

The free Green's function in cylindrical waves reads~\cite{Tsang},
\begin{equation}\label{eq:G0}
\begin{split}
&\mathbb{G}_0(\textbf{r},\mathbf{r'})=\sum_{P=M,N}\sum_{n=-\infty}^{\infty}(-1)^n\int_{-\infty}^{\infty}\frac{idk_z}{8\pi}\\
&\times\left\{
  \begin{array}{ll}
  \small{\left[\textbf{RP}_{n,k_z}(\textbf{r})\otimes\textbf{P}_{-n,-k_z}(\textbf{r}')\right],} & \hbox{\small{$r'>r$}} \\
  \small{\left[\textbf{P}_{n,k_z}(\textbf{r})\otimes\textbf{RP}_{-n,-k_z}(\textbf{r}')\right],}
& \hbox{\small{$r'<r$}}.
  \end{array}
\right.
\end{split}
\end{equation}
The following relation for propagating cylindrical waves is useful for deriving the Poynting vector,
\begin{equation}\label{useful2}
\begin{split}
&\Re\left[ir\int_0^{2\pi}d\phi\left(\textbf{RP}_{n,k_z}(\textbf{r})\times\mathbf{\nabla}\times\textbf{P}'^*_{n,k_z}(\textbf{r})T_{n,k_z}^{P'P*}\right.\right.\\
&\left.\left.+\textbf{P}_{n,k_z}(\textbf{r})\times\mathbf{\nabla}\times\textbf{RP}'^*_{n,k_z}(\textbf{r})T_{n,k_z}^{PP'}\right)\cdot\vct{e}_r\right]\\&=
4\delta_{P,P'}\textrm{Re}\left[T_{n,k_z}^{PP}\right],\\
&\Re\left[ir\int_0^{2\pi}d\phi\left(\textbf{P}_{n,k_z}(\textbf{r})\times\mathbf{\nabla}\times\textbf{P}'^*_{n,k_z}(\textbf{r})\right)\cdot\vct{e}_r\right]=4\delta_{P,P'}.
\end{split}
\end{equation}

\section{Scattering of electromagnetic waves from uniaxial cylindrical
objects}\label{B}

Consider a wave propagating in an anisotropic medium, with
dielectric permittivity tensor \eqref{tensor}
and magnetic permeability $\mu(\omega)$.

Considering the system's uniaxial symmetry we look for wave-solutions in the form of
cylindrical harmonics \eqref{cylindricalharmonics},
\begin{equation}\label{ansatz}
\begin{split}
\textbf{RM}^{un}_{n,k_z}(\textbf{r})&=\left[\frac{in}{q_Mr}J_n(q_Mr)\mathbf{e}_r-J'_n(q_Mr)\mathbf{e}_\phi\right]\\&\times e^{ik_zz+in\phi},\\
\textbf{RN}^{un}_{n,k_z}(\textbf{r})&=\frac c{\omega}
\left[ik_zJ'_n(q_Nr)\mathbf{e}_r- \frac{nk_z}{q_Nr}J_n(q_Nr)\mathbf{e}_\phi\right.\\&\left.+\gamma_N q_NJ_n(q_Nr)\mathbf{e}_z\right]e^{ik_zz+in\phi},\\
\end{split}
\end{equation}
where $\gamma_N$ is some constant
which modifies the $N$-polarized cylindrical harmonic due to uniaxial anisotropy. Importantly, we do not care about keeping the harmonics \eqref{ansatz} normalized as that does not influence calculations of $\mathbb{T}$ matrix elements in which we are interested. Note that the time dependence $\exp(-i \omega t)$
is omitted here.

The following Maxwell's equations must be satisfied,
\begin{equation}\label{maxwellequations}
\begin{split}
&-\mathbf{\nabla}\times\mathbf{\nabla}\times\textbf{RM}^{un}_{n,k_z}(\textbf{r})=\frac1{c^2}\left(\widehat\varepsilon\mu\frac{\partial^2}{\partial
t^2}\textbf{RM}^{un}_{n,k_z}(\textbf{r})\right),\\
&\mathbf{\nabla}\cdot\left[\widehat\varepsilon\textbf{RM}^{un}_{n,k_z}(\textbf{r})\right]=0,\\
\end{split}
\end{equation}
with analogous relations for
$\textbf{RN}^{un}_{n,k_z}(\textbf{r})$.

Substituting expressions \eqref{ansatz} into equations \eqref{maxwellequations},
we obtain the following unnormalized wave solutions,
$$\textbf{RM}^{un}_{n,k_z}(\textbf{r})=\left[ \frac{in}{q_Mr}J_n(q_Mr)\mathbf{e}_r-J'_n(q_Mr)\mathbf{e}_\phi\right]e^{ik_zz+in\phi},$$
\begin{equation}\label{ansatzanswer}
\begin{split}
\textbf{RN}^{un}_{n,k_z}(\textbf{r})&=\frac
c{\omega}\left[ik_zJ'_n(q_Nr)\mathbf{e}_r-\frac{nk_z}{q_Nr}J_n(q_Nr)\mathbf{e}_\phi\right.\\
&\left.+\frac{\varepsilon_r}{\varepsilon_z} q_NJ_n(q_Nr)\mathbf{e}_z\right]e^{ik_zz+in\phi},\\
\end{split}
\end{equation}
where $q_M$ and $q_N$ are the wave-vector components perpendicular to the
$z$-axis for the two solutions respectively,
\begin{equation}\label{wavevectors}
q_M=\sqrt{\varepsilon_{r}\mu k^2-k_z^2}, \quad
q_N=\sqrt{\varepsilon_{z}/\varepsilon_{r}}\sqrt{\varepsilon_{r}\mu
k^2-k_z^2}.
\end{equation}
One solution is an ordinary wave and is $M$-polarized,
whereas the other one is called an extraordinary wave and possesses
$N$-polarization~\cite{Born}.

In order to solve the scattering problem for the cylinder, we expand the electromagnetic field in cylindrical basis
\eqref{cylindricalharmonics} and \eqref{ansatzanswer}, outside and inside the cylinder respectively. The expansion coefficients for the
field inside and outside can be obtained by matching boundary
conditions at the cylinder's surface for field components tangential
to the surface.

Using the definition of the $\mathbb{T}$ matrix, we describe the scattering process
of a regular magnetic wave by the field outside the cylinder, which
is
\begin{equation}\label{magnscat1}
\textbf{E}^{M,out}_{n,k_z}=\textbf{RM}_{n,k_z}+T_{n,k_z}^{MM}\textbf{M}_{n,k_z}+T_{n,k_z}^{NM}\textbf{N}_{n,k_z}%=\textbf{RM}_{n,k_z}-\mathbb{G}_0\mathbb{T}_c\textbf{RM}_{n,k_z},
\end{equation}
and the field inside the cylinder,
\begin{equation}\label{magnscat2}
\textbf{E}^{M,in}_{n,k_z}=A_{n,k_z}^{MM}\mathbf{RM}^{un}_{n,k_z}+A_{n,k_z}^{NM}\mathbf{RN}^{un}_{n,k_z}.
\end{equation}

Analogously, for an incident electric (TM) multipole field, the
field outside the cylinder becomes
\begin{equation}\label{elecscat1}
\textbf{E}^{N,out}_{n,k_z}=\textbf{RN}_{n,k_z}+T_{n,k_z}^{MN}\textbf{M}_{n,k_z}+T_{n,k_z}^{NN}\textbf{N}_{n,k_z}%=\textbf{RN}_{n,k_z}-\mathbb{G}_0\mathbb{T}_c\textbf{RN}_{n,k_z},
\end{equation}
and the field inside the cylinder,
\begin{equation}\label{elecscat2}
\textbf{E}^{N,in}_{n,k_z}=A_{n,k_z}^{MN}\mathbf{RM}^{un}_{n,k_z}+A_{n,k_z}^{NN}\mathbf{RN}^{un}_{n,k_z}.
\end{equation}

We next derive the specific form of the $\mathbb{T}$ matrix coefficients by
matching the boundary conditions for the medium, i.e. the
continuity of $E_{\phi}$, $E_{z}$, $H_{\phi}$ and $H_{z}$ across the
cylindrical surface. Plugging the explicit form of cylindrical
harmonics \eqref{cylindricalharmonics} and \eqref{ansatzanswer} into these conditions we
obtain two sets of four linear equations for the expansion
coefficients. Using $\textbf{B}=-i(c/\omega)\mathrm{\nabla}\times
\textbf{E}$ and $\textbf{H}=\textbf{B}/\mu$ we can write the system of equations for reflection and
transmission amplitudes in case of the incident magnetic waves in
the form,
\begin{equation}\label{system1}
\mathbb{M}_{n,k_z}\left( \begin{array}{c}
 A^{MM} \\
 T^{MM} \\
 A^{NM} \\
 T^{NM} \end{array} \right)_{n,k_z}= \left( \begin{array}{c}
 \frac{c}{\omega}qJ_n(qR) \\
 J'_n(qR) \\
 0 \\
 \frac{nk_z c}{qR\omega}J_n(qR) \end{array} \right),
\end{equation}
with the matrix
\begin{widetext}
\begin{equation}\label{system11}
\mathbb{M}_{n,k_z}= \left( \begin{array}{cccc}
\frac{q_M c}{\mu\omega} J_n(q_MR) & -\frac{qc}{\omega}H_n^{(1)}(qR) & 0 & 0 \\
J_n'(q_MR) & -H_n'^{(1)}(qR) & \frac{n k_z c}{q_N R\omega} J_n(q_NR) & -\frac{n k_z c}{qR\omega} H^{(1)}_n(qR) \\
0 & 0 & \frac{\varepsilon_r}{\varepsilon_z}\frac{q_Nc}{\omega} J_n(q_NR) &-\frac{qc}{\omega} H^{(1)}_n(qR)\\
 \frac{nk_zc}{\mu q_MR\omega} J_n(q_MR) & -\frac{nk_zc}{qR\omega} H^{(1)}_n(qR) & \varepsilon_r J_n'(q_NR)& -H_n'^{(1)}(qR) \end{array} \right).
\end{equation}
\end{widetext}
For the incident electric waves the linear equations are
\begin{equation}\label{system2}
\mathbb{M}_{n,k_z} \left( \begin{array}{c}
 A^{MN} \\
 T^{MN} \\
 A^{NN} \\
 T^{NN} \end{array} \right)_{n,k_z}= \left( \begin{array}{c}
 0 \\
 \frac{nk_z c}{qR\omega}J_n(qR) \\
\frac{c}{\omega}qJ_n(qR) \\
 J'_n(qR)  \end{array} \right)
\end{equation}
The solutions to these sets of equations \eqref{system1} and
\eqref{system2} are provided in Eqs.~\eqref{TMMA}-\eqref{delta5}.

\section{Small $R$ expansion of the $\mathbb{T}$ operator of the cylinder}\label{app:an}
In order to derive Eqs.~\eqref{cylErad} and \eqref{cylMrad}, we need the expansion of the $\mathbb{T}$ operator in terms of $\omega R/c$. For a cylinder made of isotropic material with magnetic permeability $\mu(\omega)$ and dielectric permittivity $\varepsilon(\omega)$, we find for the limit  {$R\ll\{\delta,c/\omega\}$},
\begin{equation}\label{TdielectricTEE0}
T_{0,k_z}^{NN}=-\frac{i\pi}{4}(\varepsilon-1)(\widetilde{k}_z^2-1)(\omega
R/c)^2, \\
\end{equation}
\begin{equation}\label{TdielectricTMM0}
T_{0,k_z}^{MM}=-\frac{i\pi}{4}(\mu-1)(\widetilde{k}_z^2-1)(\omega
R/c)^2, \\
\end{equation}
\begin{equation}\label{TdielectricTEE1}
\begin{split}
T_{1,k_z}^{NN}=T_{-1,k_z}^{NN}&=\frac{i\pi}{4}\frac{\widetilde{k}_z^2(\mu+1)(\varepsilon-1)+(\mu-1)(\varepsilon+1)}{(\varepsilon+1)(\mu+1)}\\
 &\times(\omega
R/c)^2, \\
\end{split}
\end{equation}
\begin{equation}\label{TdielectricTMM1}
\begin{split}
T_{1,k_z}^{MM}=T_{-1,k_z}^{MM}&=\frac{i\pi}{4}\frac{\widetilde{k}_z^2(\mu-1)(\varepsilon+1)+(\mu+1)(\varepsilon-1)}{(\varepsilon+1)(\mu+1)}\\&\times (\omega
R/c)^2, \\
\end{split}
\end{equation}
\begin{equation}\label{TdielectricTME1}
\begin{split}
T_{1,k_z}^{MN}=T_{1,k_z}^{NM}=-T_{-1,k_z}^{MN}=-T_{-1,k_z}^{NM}&=\frac{i\pi}{2}\frac{(\varepsilon\mu-1)
\widetilde{k}_z}{(\varepsilon+1)(\mu+1)}\\&\times (\omega
R/c)^2, \\
\end{split}
\end{equation}
where $\widetilde{k}_z=k_z/k$. Substitution of the these forms into Eqs.~\eqref{radiationN} and \eqref{radiationM} yields Eqs.~\eqref{cylErad} and \eqref{cylMrad}.
\section{Leading term of $\mathbb{T}$ operator for $\omega/c\gg R\gg\delta$}\label{D}
For  {$\omega/c\gg R\gg\delta$}, the leading term of $\mathbb{T}$ operator is the $T^{NN}_{0,k_z}$ element which
has then the  following form,
\begin{equation}\label{condT}
\begin{split}
&\lim_{\omega/c\gg R\gg\delta}T_{0,k_z}^{NN}=\\&\frac{-\pi}{\pi+2i\gamma_E+\frac{2}{(1-\widetilde{k}_z^2)(2i+\sqrt\varepsilon
\omega R/c)}+2i\log{[\sqrt{1-\widetilde{k}_z^2}\omega R/2c]}},\\
\end{split}
\end{equation}
where $\gamma_E\approx0.577$ is the Euler-Mascheroni constant. It can be numerically shown that other
elements of $\mathbb{T}$ matrix are negligible.

\bibliographystyle{apsrev}

\end{document}